\documentclass[pra,twocolumn,preprintnumbers,superscriptaddress]{revtex4-2}

\usepackage{amsthm,amsmath,amssymb}

\usepackage{mathrsfs}
\usepackage{amsthm}
\usepackage{amsmath}
\usepackage{amssymb}
\usepackage{graphicx}
\usepackage{epstopdf}
\usepackage{url}
\usepackage{subfigure}
\usepackage{float}
\usepackage{bm}
\usepackage{ifthen}
\usepackage[usenames,dvipsnames]{color}
\usepackage{mathrsfs}
\usepackage[colorlinks=true,citecolor=blue,urlcolor=black]{hyperref}
\usepackage{float}

\newcommand{\be}{\begin{equation}}
	\newcommand{\ee}{\end{equation}}

\newcommand{\sket}[1]{{\ensuremath{\lvert#1\rangle}}}
\newcommand{\lket}[1]{{\ensuremath{\left\lvert#1\right\rangle}}}
\newcommand{\ket}[1]{\if@display\lket{#1}\else\sket{#1}\fi}

\newcommand{\sbra}[1]{{\ensuremath{\langle#1\rvert}}}
\newcommand{\lbra}[1]{{\ensuremath{\left\langle#1\right\rvert}}}
\newcommand{\bra}[1]{\if@display\lbra{#1}\else\sbra{#1}\fi}

\newcommand{\sbraket}[2]{{\ensuremath{\langle#1\rvert#2\rangle}}}
\newcommand{\lbraket}[2]{{\ensuremath{\left\langle#1\!\left\rvert\vphantom{#1}#2\right.\!\right\rangle}}}
\newcommand{\braket}[2]{\if@display\lbraket{#1}{#2}\else\sbraket{#1}{#2}\fi}

\newcommand{\sketbra}[2]{{\ensuremath{\lvert #1\rangle\!\langle #2\rvert}}}
\newcommand{\lketbra}[2]{{\ensuremath{\left\lvert #1\right\rangle\!\!\left\langle #2\right\rvert}}}
\newcommand{\ketbra}[2]{\if@display\lketbra{#1}{#2}\else\sketbra{#1}{#2}\fi}

\newcommand{\cX}{\mathcal{X}}

\theoremstyle{plain}

\theoremstyle{definition}

\begin{document}

\title{Experimental quantum secure network with digital signatures and encryption}
	
\author{Hua-Lei Yin}\email{hlyin@nju.edu.cn}
\affiliation{National Laboratory of Solid State Microstructures, School of Physics, Collaborative Innovation Center of Advanced Microstructures, Nanjing University, Nanjing 210093, China}
\author{Yao Fu}\email{yfu@iphy.ac.cn}
\affiliation{Beijing National Laboratory for Condensed Matter Physics and Institute of Physics, Chinese Academy of Sciences, Beijing 100190, China}
\affiliation{MatricTime Digital Technology Co. Ltd., Nanjing 211899, China}
\author{Chen-Long Li}
\author{Chen-Xun Weng}
\author{Bing-Hong Li}
\author{Jie Gu}
\author{Yu-Shuo Lu}
\author{Shan Huang}
\affiliation{National Laboratory of Solid State Microstructures, School of Physics, Collaborative Innovation Center of Advanced Microstructures, Nanjing University, Nanjing 210093, China}
\author{Zeng-Bing Chen}\email{zbchen@nju.edu.cn}
\affiliation{National Laboratory of Solid State Microstructures, School of Physics, Collaborative Innovation Center of Advanced Microstructures, Nanjing University, Nanjing 210093, China}
\affiliation{MatricTime Digital Technology Co. Ltd., Nanjing 211899, China}		
\date{\today}

\begin{abstract}
Cryptography promises four information security objectives, namely, confidentiality, integrity, authenticity, and non-repudiation, to support trillions of transactions annually in the digital economy. Efficient digital signatures, ensuring the integrity, authenticity, and non-repudiation of data with information-theoretical security are highly urgent and intractable open problems in cryptography. Here, we propose a protocol of high-efficiency quantum digital signatures using secret sharing, one-time universal$_2$ hashing, and the one-time pad. We just need to use a 384-bit key to sign documents of up to $2^{64}$ lengths with a security bound of $10^{-19}$. If one-megabit document is signed, the signature efficiency is improved by more than $10^8$ times compared with previous quantum digital signature protocols. Furthermore, we build the first all-in-one quantum secure network integrating information-theoretically secure communication, digital signatures, secret sharing, and conference key agreement and experimentally demonstrate this signature efficiency advantage. Our work completes the cryptography toolbox of the four information security objectives.
\end{abstract}

\maketitle


\section{INTRODUCTION}\label{sec1}

Fast developing driverless, blockchain and artificial intelligence technologies, as well as digital currency systems, will soon require a more robust network with security against quantum attacks~\cite{fedorov2018quantum}. A promising blueprint for such a network ensures hash functions, encryption algorithms, and digital signatures with information-theoretical security, which cannot be met in the current Internet with public-key infrastructure~\cite{menezes2018handbook}. Currently, widely implemented one-way hash functions, such as Message Digest-5~\cite{wang2005break} and Secure Hash Algorithm-1~\cite{wang2005finding}, are no longer secure.
For example, since 2017, one can utilize two different files to obtain the identical hash value after conducting  Secure Hash Algorithm-1~\cite{Stevens:2017:The}. Additionally, in 2020, the public-key encryption algorithms~\cite{kleinjung2010factorization,kleinjung2017computation,boudot2020comparing}, based on the computational complexity of factorization and discrete logarithm, have both been compromised at the 795-bit level~\cite{boudot2020comparing}. More seriously, quantum computers can in principle attack public-key cryptosystems with any number of bits~\cite{shor1994algorithms}.

Unlike public-key cryptography, one-time pad (OTP) encryption based on a symmetric key allows a message to be transmitted with information-theoretical confidentiality~\cite{shannon1949communication} over a standard communication channel, with the symmetric key being securely established using quantum key distribution~\cite{bennett2014quantum} and the attackers' computational power being unrestricted. Currently, there are several experimental demonstrations and commercial applications of quantum key distribution around the world~\cite{chen2021integrated,liu2020drone,kwek2021chip,zhang2020design,guo2021toward}.

\begin{figure}[ht!]
	\begin{center}
		\includegraphics[width=8.5cm]{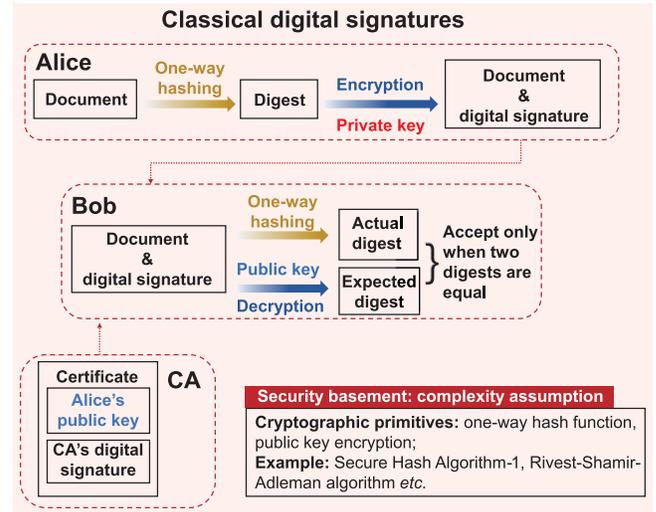}
		\caption{Schematic diagram of classical digital signatures. Alice uses a private key to encrypt the digest to obtain the signature, where the digest is acquired via a fixed one-way hash function. She sends the document along with the signature to Bob. Bob utilizes the same one-way hash function and the corresponding public key to acquire two digests. Then, he only accepts the signature if the two digests are identical. Thereinto, a digital certificate issued by the certificate authority (CA) guaranties the validity of the public key. Here, we omit the generation process of the private and public keys.}
		\label{f1}
	\end{center}
\end{figure}

Quantum key distribution~\cite{bennett2014quantum} and quantum secure direct communication~\cite{long2002theoretically,qi202115,Sheng:2022:One} only ensure confidentiality, which is, however, an incomplete solution to the remaining cryptographic tasks. Three other fundamental information security objectives are integrity, authenticity, and non-repudiation~\cite{menezes2018handbook}. These other tasks are usually realized in classical cryptosystems by digital signatures using one-way hash functions and public-key encryption algorithms, as shown in Fig.~\ref{f1}. Digital signatures~\cite{menezes2018handbook} play a vital role in software distribution, e-mails, web browsing, and financial transactions, but they become insecure as one-way hash functions and public-key encryption algorithms used therein are breakable by either classical or quantum computers~\cite{lyubashevsky2021lattice}.

Unlike classical solutions~\cite{menezes2018handbook}, quantum digital signatures (QDS) use quantum laws to sign a document with information-theoretical integrity, authenticity, and non-repudiation. In 2001, the first rudiment of the QDS was introduced~\cite{gottesman2001quantum}, but it could not be implemented. Developments in the last decade have removed impractical requirements of the QDS, such as high-dimensional single-photon state~\cite{clarke2012experimental}, quantum memory~\cite{dunjko2014quantum}, and secure quantum channels~\cite{Yin:practical:2016,Amiri:Secure:2016,lu2021efficient,weng2021secure}, enabling demonstrations of the QDS in various experimental systems~\cite{Yin:2017:Experimental,Yin:2017:Exp,roberts2017experimental,an2019practical,Richter2021Agile}. However, the resulting schemes still have serious limitations that require an approximately $10^{5}$-bit key to sign only a one-bit document. For a gigahertz system, the best signature rate reported thus far is less than 1 time per second (tps) for a one-bit at a 100-km transmission distance~\cite{an2019practical}. Additionally, it is unknown how to efficiently sign multi-bit documents with information-theoretical security~\cite{wang2015security}, which makes all known single-bit-type QDS protocols far from practical applications~\cite{gottesman2001quantum,clarke2012experimental,dunjko2014quantum,Yin:practical:2016,Amiri:Secure:2016,lu2021efficient,weng2021secure,Yin:2017:Experimental,Yin:2017:Exp,roberts2017experimental,an2019practical,Richter2021Agile}. Thus, a high-efficiency QDS that is as feasible as quantum private communication (using quantum key distribution)~\cite{chen2021integrated} is highly desirable and remains an unsolved open problem. Note that a probabilistic one-time delegation of signature authority protocol was proposed and demonstrated using entanglement correlation~\cite{roehsner2021probabilistic}.

Here, we propose a one-time universal$_2$ hashing (OTUH)-QDS protocol capable of signing an arbitrarily long document with information-theoretical security. For example, just with a 384-bit key, our protocol can sign documents of up to $2^{64}$ lengths with a security bound of $10^{-19}$. Furthermore, we propose, for the first time, the concept of OTUH: a completely random and different universal$_2$ hash function~\cite{carter1979universal} used for each digital signature. Our protocol not only uses OTUH and OTP as the underlying cryptography layer but also uses secret sharing to realize the perfect bits correlation of the three parties, and then build an asymmetric key relationship for Alice and Bob. Secret sharing can be implemented with information-theoretical security using quantum secret sharing, quantum key distribution, or future quantum internet with solid-state entanglement. Additionally, we simulate the performances of our OTUH-QDS protocol based on various quantum communication protocols. The simulation results show that for a gigahertz system, the signature rates are more than $10^{4}$ tps in the metropolitan area, which represents an efficiency improvement of at least eight orders of magnitude for signing a one-megabit document. Additionally, we experimentally construct a quantum secure network to realistically demonstrate cryptographic primitives~\cite{menezes2018handbook} with information-theoretical security, such as private communication, digital signatures, secret sharing, and conference key agreement. In our experiment, the signature efficiency can be achieved $1.43\times10^{8}$ times that of the previous work in Ref.~\cite{Amiri:Secure:2016} considering the improvements in signature rates for signing a 130,250-byte document over 101-km fiber, and the security bound is as small as $10^{-32}$, which shows a significant advantage.

\section{RESULTS}

\noindent
\textbf{Efficient QDS protocol}\label{sec2}

\begin{figure*}[ht!]
	\begin{center}
		\includegraphics[width=16cm]{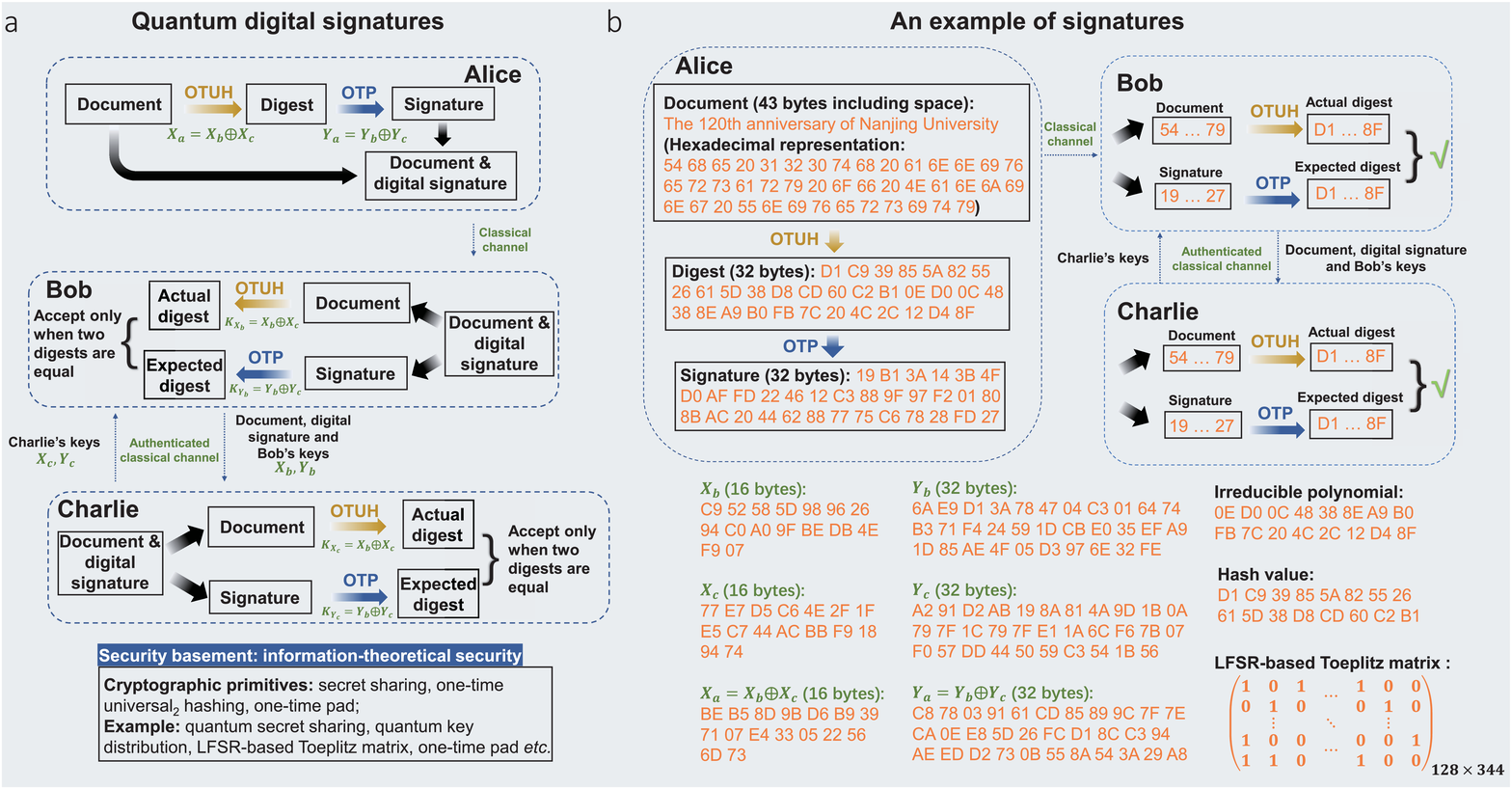}
		\caption{Schematic diagram for the QDS and the corresponding example. (a) Compared with the classical scheme, Charlie plays the role of a certificate authority. Alice's key can be viewed as a quantum private key, while Bob's key is a quantum public key; in our protocol, they are asymmetric. The information-theoretically secure OTUH replaces the fixed one-way hash function. Here, we omit the pre-distribution stage for information-theoretically secure asymmetric quantum key generation, which replaces the classical private and public key generation process. (b) As an example, we sign a document of ``The 120th anniversary of Nanjing University." The details of the document, digest, signature, irreducible polynomial, and key bit strings are shown in hexadecimal.}
		\label{f2}
	\end{center}
\end{figure*}

Before executing our OTUH-QDS protocol, three parties, Alice, Bob, and Charlie, will perform the pre-distribution stage, which is analogous to the private and public keys generation procedure in classical digital signatures. Alice, Bob, and Charlie each have two key bit strings $\{X_{a},~X_{b},~X_{c}\}$ with $n$ bits and $\{Y_{a},~Y_{b},~Y_{c}\}$ with $2n$ bits, where the key bit strings meet the perfect correlation $X_{a}=X_{b}\oplus X_{c}$ and $Y_{a}=Y_{b}\oplus Y_{c}$, respectively. The pre-distribution stage can be realized using quantum communication protocols (see Methods), such as quantum key distribution~\cite{wang2018TFlargemisalignment,ma2018phase,lin2019asymptotic,Lodewyck2007Quantum,lo2012measurement} and quantum secret sharing~\cite{hillery1999quantum,fu2015longdistance,gu2021secure,schmid2005experimental,gu2021differential}. Before executing the signature, Alice is the signer, and both Bob and Charlie can be the receiver because of the symmetry between Bob and Charlie. Here, we suppose that Alice signs an $m$-bit document, denoted by $Doc$, to the desired recipient, Bob. Therefore, Bob is the specified receiver, and Charlie automatically becomes the verifier. Our proposed approach utilizes secret sharing, OTUH, and OTP to generate and verify signatures, as shown in Fig.~\ref{f2}a. We remark that the keys of signer, Alice, and receiver, Bob, are asymmetric because $X_{a}\neq X_{b}$ and $Y_{a}\neq Y_{b}$. After completing the pre-distribution stage, the three parties can implement the signature stage at any time.

(i) \emph{Signing of Alice}---First, Alice uses a local quantum random number, which can be characterized by an $n$-bit string $p_a$ (see Supplementary data), to randomly generate an irreducible polynomial $p(x)$ of degree $n$~\cite{menezes2018handbook}. Second, she uses the initial vector (key bit string $X_{a}$) and irreducible polynomial (quantum random number $p_{a}$) to generate a random linear feedback shift register-based (LFSR-based) Toeplitz matrix~\cite{krawczyk1994lfsr} $H_{nm}$, with $n$ rows and $m$ columns. Third, she uses a hash operation with $Hash$= $H_{nm} \cdot Doc$ to acquire an $n$-bit hash value of the $m$-bit document. Fourth, she exploits the hash value and the irreducible polynomial to constitute the $2n$-bit digest $Dig=(Hash||p_a)$. Fifth, she encrypts the digest with her key bit string $Y_{a}$ to obtain the $2n$-bit signature $Sig=Dig\oplus Y_{a}$ using OTP. Finally, she uses the public channel to send the signature and document $\{Sig,~Doc\}$ to Bob.

(ii) \emph{Verification of Bob}---Bob uses the authentication classical channel to transmit the received $\{Sig,~Doc\}$, as well as his key bit strings $\{X_{b},~Y_{b}\}$, to Charlie. Then, Charlie uses the same authentication channel to forward his key bit strings $\{X_{c},~Y_{c}\}$ to Bob. Bob obtains two new key bit strings $\{K_{X_{b}}=X_{b}\oplus X_{c},~K_{Y_{b}}=Y_{b}\oplus Y_{c}\}$ by the XOR operation. Bob exploits $K_{Y_{b}}$ to obtain an expected digest and bit string $p_b$ via XOR decryption. Bob utilizes the initial vector $K_{X_{b}}$ and irreducible polynomial $p_b$ to establish an LFSR-based Toeplitz matrix. Bob uses a hash operation to acquire an $n$-bit hash value and then constitutes a $2n$-bit actual digest. Bob will accept the signature if the actual digest is equal to the expected digest. Then, he informs Charlie of the result. Otherwise, Bob rejects the signature and announces to abort the protocol.

(iii) \emph{Verification of Charlie}---If Bob announces that he accepts the signature, Charlie then uses his original key and the key sent to Bob to create two new key bit strings $\{K_{X_{c}}=X_{b}\oplus X_{c},~K_{Y_{c}}=Y_{b}\oplus Y_{c}\}$. Charlie employs $K_{Y_{c}}$ to acquire an expected digest and bit string $p_c$ via XOR decryption. Charlie uses a hash operation to obtain an $n$-bit hash value and then constitutes a $2n$-bit actual digest, where the hash function is an LFSR-based Toeplitz matrix generated by initial vector $K_{X_{c}}$ and irreducible polynomial $p_c$. Charlie accepts the signature if the two digests are identical. Otherwise, Charlie rejects the signature.

To show more clearly how our protocol works, Fig.~\ref{f2}b shows an example of signing a document ``The 120th anniversary of Nanjing University."

\bigskip
\noindent
\textbf{Security proof}\label{SA}

In a QDS scheme, either Alice or Bob can be the attacker. Thus, Alice and Bob distrust each other, whereas the verifier, Charlie, is always trusted. Bob and Charlie will cooperate to counter Alice's repudiation attack. Alice and Charlie will collaborate to counter Bob's forgery attack. Besides, we also consider the robustness of our protocol.

\emph{Security against forgery.}
When Charlie accepts the tampered document forwarded by Bob, Bob's forgery attack is considered successful. There are two cases of Bob's forgery attack. First, Bob can generate a new document and signature if Alice has not signed a document at all. Second, Bob can change the document and signature if Alice has signed the document. According to our protocol, Charlie accepts the signature if and only if he obtains the identical digest by decrypting the signature with OTP and hashing the document with OTUH, respectively. Note that before Bob forwards the document, signature, and his key bit strings to Charlie, Bob cannot obtain the key bit strings of Charlie. In the first case, Bob has no information since Alice did not send any information. The only thing Bob can do is correctly guess Alice's key bit strings $X_{a}$ and $Y_{a}$, i.e., guessing Charlie's key bit strings $X_{c}$ and $Y_{c}$ based on $X_{a}=X_{b}\oplus X_{c}$ and $Y_{a}=Y_{b}\oplus Y_{c}$. The probability of guessing correctly is at most $1/2^{n}$ since Bob has no information of key bit strings $X_{a}$ and $Y_{a}$ with $n$ and $2n$ bits, respectively. In the second case, Bob also has no information on the universal$_2$ hash function (initial vector $X_{a}$ and irreducible polynomial $p_{a}$ for the LFSR-Based Toeplitz matrix) used by Alice since the digest has been encrypted to a signature using OTP. Besides, Bob cannot obtain any information from the previous signing round because their keys are refreshed and the corresponding universal$_2$ hash function is updated in each round in our protocol. Compared to guessing the key bit strings of Alice or Charlie, Bob's best strategy is to guess the irreducible polynomial $p_{a}$ of the LFSR-Based Toeplitz matrix. The collision probability of universal$_2$ hashing by the LFSR-based Toeplitz matrix can be determined by $m/2^{n-1}$ (see Methods), which implies that one can find two distinct documents with identical hash values by randomly guessing the irreducible polynomial $p_{a}$. Therefore, for any case, the probability of a successful forgery can be bounded by
\begin{equation}
	\epsilon_{\rm for}=\frac{m}{2^{n-1}},
\end{equation}
where $m$ is the length of the document $Doc$ and $n$ is the order of the irreducible polynomial $p_{a}$.

Note that our proof is information-theoretically secure, even though Bob has unlimited computing power. We emphasize the importance of our proposed OTUH, where the universal$_2$ hash function is only used once and then updated. Bob cannot obtain any information from the previously signed round because their keys and irreducible polynomial are refreshed in every round. Bob cannot do anything at all apart from randomly guessing. Moreover, before Bob sends the document and signature to Charlie, Bob cannot be sure if he guessed correctly even if he exhausts all the possibilities. Bob's forgery attack in our OTUH-QDS protocol is successfully related to Eve's attack in information-theoretically secure message authentication~\cite{krawczyk1994lfsr,fung2010practical} (details can be found in Supplementary data).

\emph{Security against repudiation.}
Successful repudiation means that Alice makes Bob accept the signature, while Charlie rejects it. For Alice's repudiation attacks, Bob and Charlie are both honest and trust each other. Note that Bob and Charlie must forward their key bit strings to each other through an authenticated classical channel. The authenticated channel used ensures that Alice knows about the transmitted information between Bob and Charlie but cannot tamper with it. Then, Bob and Charlie can recover the identical key bit strings through the XOR operation $K_{X_{b}}=X_{b}\oplus X_{c}=K_{X_{c}}$ and $K_{Y_{b}}=Y_{b}\oplus Y_{c}=K_{Y_{c}}$. Bob and Charlie obtain the same irreducible polynomial $p_{b}=p_{c}$ through OTP decryption. They will make the same decision for the same document, signature, key bit strings, and irreducible polynomial. Therefore, our QDS protocol is naturally immune to repudiation. The probability of repudiation is zero when we ignore the insignificant failure probability of secure message authentication.

Note that in all known QDS protocols, the symmetry between Bob and Charlie is used to counter Alice's repudiation attacks. Compared to partial symmetry in previous protocols, Bob and Charlie will have identical key bit strings in our protocol after performing the QDS process. In addition, there is no help for Alice's repudiation attack, even though she is dishonest in the pre-distribution stage because we allow Alice to obtain all information from Bob (Charlie) about $X_{b(c)}$ and $Y_{b(c)}$.

\begin{figure*}[ht!]
	\begin{center}
		\includegraphics[width=16cm]{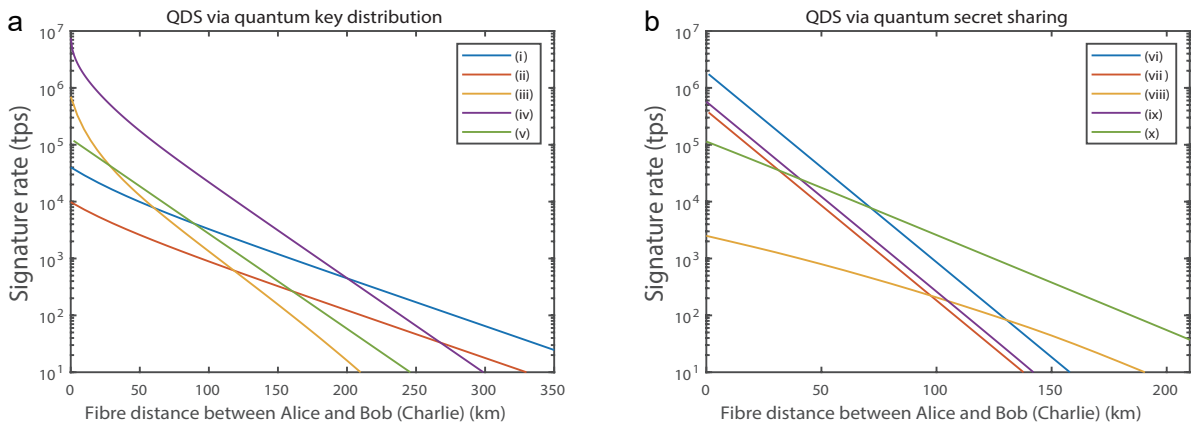}
		\caption{Signature rates versus fiber distances. (a) The rate via quantum key distribution. (b) The rate via quantum secret sharing. Lines labeled (i)--(v) represent quantum key distribution protocols: (i) sending-or-not-sending twin-field~\cite{wang2018TFlargemisalignment}, (ii) phase-matching~\cite{ma2018phase}, (iii) discrete-modulated continuous-variable~\cite{lin2019asymptotic}, (iv) Gaussian-modulated continuous-variable~\cite{Lodewyck2007Quantum}, and (v) measurement-device-independent~\cite{lo2012measurement}. Lines labeled (vi)--(x) represent quantum secret sharing protocols: (vi) prepare-and-measure~\cite{hillery1999quantum},  (vii) measurement-device-independent~\cite{fu2015longdistance}, (viii) round-robin~\cite{gu2021secure}, (ix) single-qubit~\cite{schmid2005experimental} and (x) differential-phase-shift~\cite{gu2021differential}. To simplify, let the channel loss of Alice--Bob and Alice--Charlie be the same. We need a 384-bit key for performing each digital signature with a security bound of approximately $1.1\times10^{-19}$ for documents of up to $2^{64}$ length. For a gigahertz system, the signature rates are more than $10^{4}$ tps in the metropolitan area.}
		\label{simuandexp}
	\end{center}
\end{figure*}

\emph{Robustness.}
The robustness quantifies the probability that Bob rejects the signature when the three parties are truthful. If Alice, Bob, and Charlie are all truthful, there are the relations of irreducible polynomial $p_a=p_b=p_c$ and key bit strings $X_a=K_{X_b}=K_{X_c}$ and $Y_a=K_{Y_b}=K_{Y_c}$. Thus, they will use the same universal$_{2}$ hash function and generate the same actual digest. The signature will be accepted naturally. The probability of honest aborting is zero, though, in the pre-distribution stage, we ignore the insignificant failure probability of classical bit error-correction of quantum communication protocols.

Note that the verification step of error correction in quantum key distribution and quantum secret sharing is usually realized using the universal$_{2}$ hashing and OTP, which is related to the information-theoretically secure message authentication~\cite{fung2010practical}. The verification step ensures that the classical bit error correction is successful with a small failure probability.

Thus, if one uses a 128-bit and 256-bit keys for OTUH and OTP, respectively, the security bound of our OTUH-QDS is less than $2^{64}/(2^{128-1})\approx1.1\times10^{-19}$ for documents of up to $2^{64}$ lengths.

\bigskip
\noindent
\textbf{Simulation results of the QDS}\label{sec3}

Secret sharing in the pre-distribution stage allows the key bit strings of Alice, Bob, and Charlie to satisfy $X_{a}=X_{b}\oplus X_{c}$ and $Y_{a}=Y_{b}\oplus Y_{c}$, which can be implemented with information-theoretical security using quantum secret sharing, quantum key distribution, or future quantum internet with solid-state entanglement. Meanwhile, a full-blown quantum internet, with functional quantum computers and quantum repeaters as nodes connected through quantum channels, is being developed. The first prototype of the quantum internet has been realized with remote solid-state qubits~\cite{Pompili2021network} in multiparty entanglements applicable to secret sharing. To date, there is no workable quantum-secure asymmetric cryptosystem. With the help of secret sharing, our framework represents the first practical quantum asymmetric cryptosystem ($X_{a}\neq X_{b}$ and $Y_{a}\neq Y_{b}$) immediately applicable to secure digital signatures.

\begin{figure*}[ht!]
	\begin{center}
		\includegraphics[width=16cm]{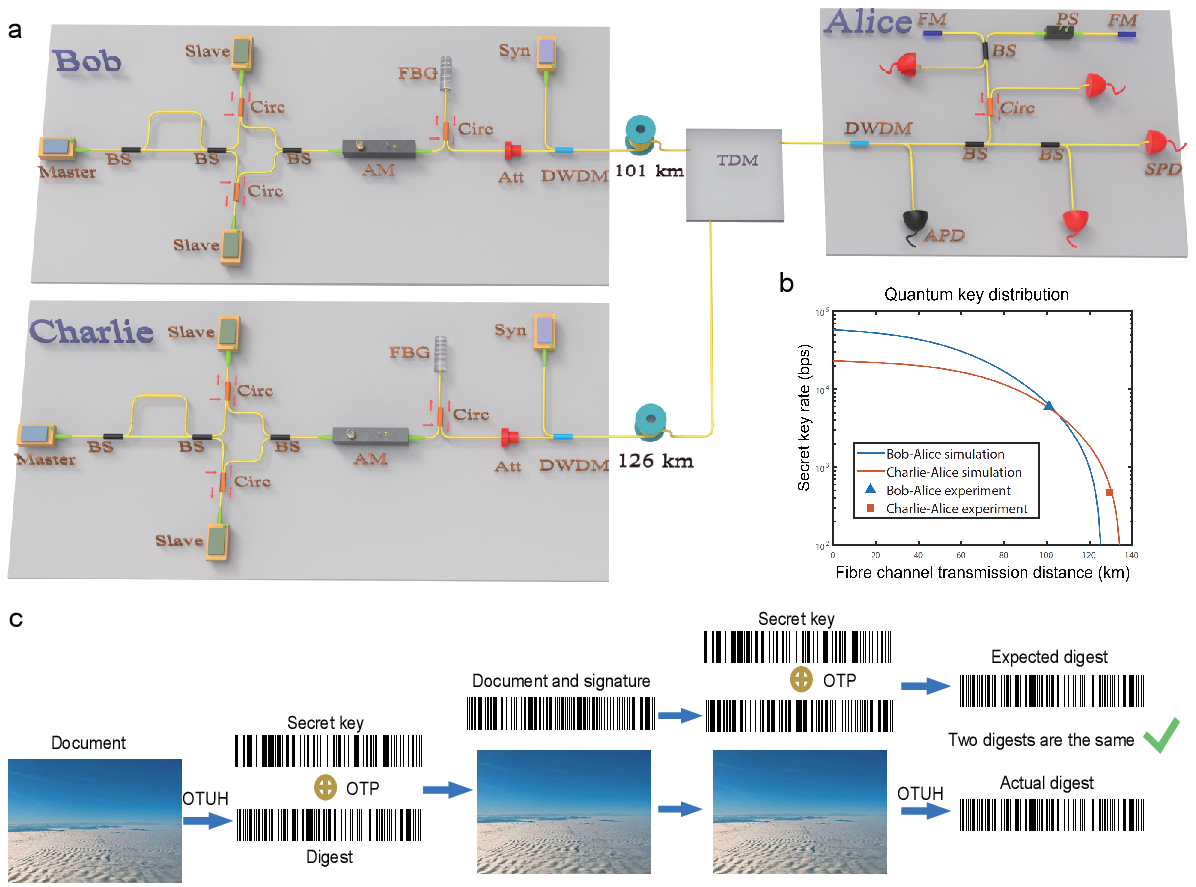}
		\caption{Experimental setup of the quantum secure network. (a) Bob (Charlie) exploits a master laser, an asymmetric interferometer, two slave lasers, two circulators (Circ), and a beam splitter (BS) to prepare optical pulses in the Z and X bases by controlling the trigger electrical signal of slave lasers. The decoy signals are generated by the amplitude modulator (AM), whereas the vacuum state is produced by removing the triggered signal of slave lasers. The optical pulses pass through a set of fiber Bragg grating (FBG), circulator, and attenuator (Att) to be modulated at the single-photon level. The synchronization (Syn) signal is transmitted to quantum channels with a dense wavelength division multiplexer (DWDM). The synchronization pulse is detected by an avalanche photodiode (APD). A biased beam splitter is utilized to realize a passive basis measurement with a single-photon detector (SPD). An asymmetric interferometer is formed by two Faraday mirrors (FM), a phase shifter (PS), and a beam splitter. The quantum signals sent by Bob (Charlie) are received by Alice by time-division multiplexing (TDM). (b) Experimental results of the decoy-state quantum key distribution. The blue and red curves of the secret key rate correspond to the simulation results using experimental parameters. (c) Demonstration of quantum digital signatures. The document to be signed with a length of 130,250 bytes includes the timestamp, identity number of the desert image, and the image itself.
		} \label{f4}
	\end{center}
\end{figure*}

Here, we simulate the performances of our OTUH-QDS protocol using various quantum key distribution~\cite{wang2018TFlargemisalignment,ma2018phase,lin2019asymptotic,Lodewyck2007Quantum,lo2012measurement} and quantum secret sharing~\cite{hillery1999quantum,fu2015longdistance,gu2021secure,schmid2005experimental,gu2021differential} protocols (see Supplementary data), as depicted in Fig.~\ref{simuandexp}. For a gigahertz system and the symmetric channel case, the simulation results show that if the fiber distance between Alice and Bob (or Charlie) is less than 50 km, one can implement digital signatures up to $10^{4}$ tps, even for the $2^{64}$-bit document. Therefore, one can conduct tens of thousands of transactions per second secured by digital signatures in the metropolitan area network~\cite{chen2021integrated}.

Our OTUH-QDS protocol has two significant features. First, as the length of the signed document increases up to $10^{19}$ bits, the key bits consumed by our protocol are almost constant, while having sufficient security as discussed above. This means that the signature efficiency of our protocol has a significant advantage over previous single-bit-type QDS protocols~\cite{gottesman2001quantum,clarke2012experimental,dunjko2014quantum,Yin:practical:2016,Amiri:Secure:2016,lu2021efficient,weng2021secure,Yin:2017:Experimental,Yin:2017:Exp,roberts2017experimental,an2019practical,Richter2021Agile}. As the length of the document is increased from $1$ to $10^{19}$ bits, our quantum resource consumption (384-bit key) does not change, which means that the signature time will not increase. Therefore, our protocol has the signature efficiency advantage from $10^{2}$ to $10^{21}$, compared to previous protocols that need at least $10^5$ bits to sign one bit~\cite{Yin:2017:Experimental,Yin:2017:Exp,roberts2017experimental,an2019practical,Richter2021Agile}. Second, our OTUH-QDS protocol is flexible for all applications. In our QDS protocol, all known and future developed quantum secret sharing and quantum key distribution protocols can be used for the perfect bits correlation of the three parties (secret sharing). Additionally, the universal$_2$ hash function should not be restricted to the LFSR-based Toeplitz matrix~\cite{krawczyk1994lfsr}, which is used here just as an example.

\begin{table*}[t]
	\centering
	\caption{List of the experimental results of the QDS in Ref.~\cite{Amiri:Secure:2016} and our OTUH-QDS protocol. At each time, a document of $10^6$ bits is assumed to be signed.}
	\setlength{\tabcolsep}{4mm}{
	\begin{tabular}{c|c|c}
			\hline
			\hline
			 &  Ref.~\cite{Amiri:Secure:2016}&  our protocol  \\
			\hline
			distance between Bob and Charlie (km)& 101+126=227 & 101+126=227  \\
			keys consumption (bit) &  $1.09\times10^{12}~(4.66\times 10^{11})$ & 384   \\
			valid keys length per second (bit) & 9314  & 470 \\
			signature rate (tps) &  $8.54\times10^{-9}~(2.00\times 10^{-8})$ & 1.22  \\
			security bound ($\epsilon$) &  $10^{-32}~(10^{-10})$ & $10^{-32}$  \\
			\hline
			\hline
	\end{tabular}}
	\label{qds}
\end{table*}

\bigskip
\noindent
\textbf{Experimental results of the QDS}

To verify the efficiency and feasibility of our OTUH-QDS protocol, we established a three-node quantum secure network containing two end nodes (Bob and Charlie) and an intermediate node (Alice), as shown in Fig.~\ref{f4}a. Two point-to-point quantum key distribution links are built between Alice--Bob and Alice--Charlie using the decoy-state protocol with a time-bin phase encoding system~\cite{yin2020experimental}. Bob (Charlie) multiplexes the 1570-nm synchronization pulse with a 1550-nm quantum signal by a dense wavelength division multiplexer, transmitted through a 101-km (126-km) single-mode optical fiber to Alice; the corresponding loss of quantum channels is 19 (24.3) dB, and the system clock frequency is 200 MHz. To reduce the insertion loss of the receiving end, we take advantage of time-division multiplexing by manually switching fiber links. A classic network is used to communicate in the postprocessing stage, including parameter estimation, error correction, and privacy amplification (details can be found in the Supplementary data).

In Fig.~\ref{f4}b, the blue (red) symbol refers to the experimental secret key rates of quantum key distribution between Bob--Alice (Charlie--Alice) with 6021 (470) bits per second by considering the finite-size effects, fitting well with our simulation curves. The blue and red curves are both flattened in the short distance since we introduce dead times of 10 and 25 $\mu$s for the gated-mode InGaAs/InP single-photon detector, respectively.

Here, we describe the experimental demonstration of quantum digital signatures for a 130,250-byte ($1.042\times10^{6}$-bits) document over 101-km fiber, as shown in Fig.~\ref{f4}c. The secret sharing is realized so that Alice performs an XOR operation for her two key bit strings. One key bit string is shared with Bob using the time-bin phase encoding quantum key distribution, and the other is shared with Charlie by exploiting another quantum key distribution system. The signed document includes the timestamp, identity number of the desert image, and the image itself. The digest is composed of the 128-bit hash value generated through OTUH and the 128-bit irreducible polynomial, and then, it is encrypted to form a signature by OTP. Both the digest and signature are displayed as bar codes and have the same size of 32 bytes (256 bits). The actual and expected digests are identical, indicating that we have applied successful quantum digital signatures with information-theoretical security.

For a fair comparison, we also demonstrate the single-bit-type QDS of Ref.~\cite{Amiri:Secure:2016} using the same experimental system. Table~\ref{qds} shows the results. For signing a single-bit document, the length of the raw key using the method in Ref.~\cite{Amiri:Secure:2016} (without error correction and privacy amplification) is $2.88\times10^{6}$ bits. For a multi-bit document, for example, one megabit, at least the length of the key with $2.88\times10^{12}$ bits is required~\cite{wang2015security,Yin:2017:Experimental}. Therefore, the signature rate of our OTUH-QDS protocol is $1.22$ tps, whereas using the method of Ref.~\cite{Amiri:Secure:2016}, it is only $3.23\times10^{-9}$ tps if we let the size of each signed document be $10^{6}$ bits. Fig.~\ref{f4}c depicts the experimental demonstration of the QDS. We only require less than one second to run the quantum secure network, whereas using the method of Ref.~\cite{Amiri:Secure:2016}, it will take approximately as long as four years to accumulate data.

\begin{figure*}[t]
	\begin{center}
		\includegraphics[width=16cm]{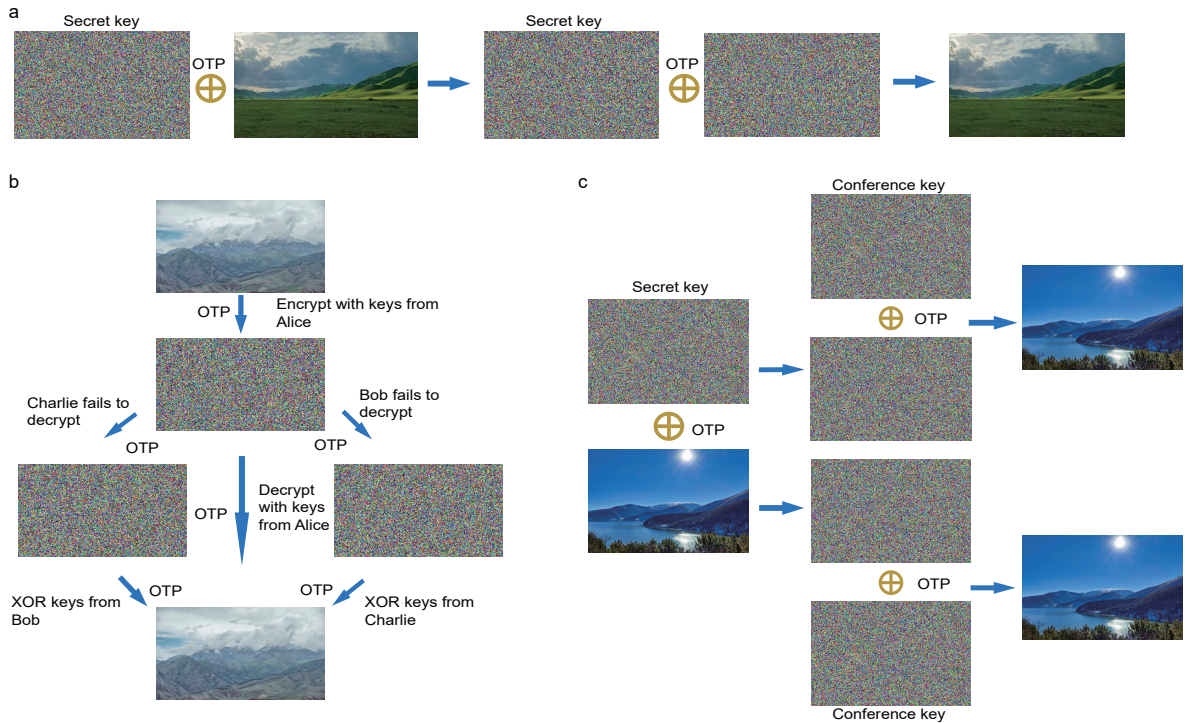}
		\caption{Experimental demonstration of other cryptographic tasks. All encrypted images and secret (conference) keys are demonstrated as images of white noise. (a) Encryption. A prairie image with a size of 112,500 bytes is encrypted via an OTP, utilizing identical secret keys shared between Bob and Charlie, to realize the perfectly private communication. (b) Secret sharing. An image of a mountain with a size of 79,800 bytes is used to realize provable secret sharing. Bob and Charlie can decrypt the image only when they work together to reconstruct the secret keys of Alice. (c) Conference key agreement. An image of a lake with a size of 139,500 bytes is adopted to implement group encryption. All users of this group can obtain the information of the encrypted figure separately.}
		\label{demonstration}
	\end{center}
\end{figure*}

We would like to clarify two main reasons why our protocol shows a huge improvement in the signature efficiency compared with early QDS schemes. First, the early QDS protocols set the threshold value and compare it with the mismatch rate of bit strings that are from the other two parties to determine whether to accept or not. However, after error correction and privacy amplification in our pre-distribution stage, the secret keys of the three parties are perfectly correlated, which satisfies the relationship $X_a=X_b\oplus X_c $ and $Y_a=Y_b\oplus Y_c$. Besides, Bob and Charlie have identical key bit strings instead of partial symmetric key bit strings in previous protocols. This change in key bit strings will result in approximately two orders of magnitude improvement in the signature efficiency due to removing the reception threshold inequality and the corresponding statistical fluctuations. Second, we use the universal$_2$ hash function to implement uniform mapping of long documents to short hash values with information-theoretical security. Any attempt to change the document will change the hash value with probability $1-\epsilon_{\rm for}$. Since the hash value and universal$_2$ hash function are completely unknown (encrypted by a one-time pad), one cannot do anything but randomly guess them. Therefore, we can use a fixed key length to sign documents of almost any length. However, the core of the previous QDS solutions is for the signer to sign document with bit by bit~\cite{wang2015security}, which means that one needs at least $m$ times the key length to sign an $m$-bit document. Moreover, in previous studies, information-theoretical security has not been proven for signing multi-bit documents. However, in our experiment, it will result in at least six orders of magnitude improvement in the signature efficiency for the $1.042\times10^{6}$-bit-signed document.

\bigskip
\noindent
\textbf{Demonstration of other cryptographic tasks}

To demonstrate the full-function information security objectives, shown in Fig.~\ref{demonstration}, we demonstrate the other three cryptographic tasks in our quantum secure network with information-theoretical security, including encryption, secret sharing, and conference key agreement. Fig.~\ref{demonstration}a illustrates quantum private communication with Alice's help as a trust relay. Alice performs an XOR operation for her two key bit strings that are shared with Bob and Charlie. To realize secure encryption between Bob and Charlie, Alice announces the XOR result as a key relay to make Bob and Charlie share identical keys. To realize quantum communication, a prairie image with 112,500 bytes is encrypted via OTP.

In the secret-sharing task, Alice is an honest dealer, while Bob and Charlie are the players. Therefore, either Bob or Charlie is a dishonest player, which can be ensured in quantum key distribution links Charlie--Alice (Bob is the attacker) and Bob--Alice (Charlie is the attacker), respectively. Before the implementation of secure secret sharing, only Alice knows that the XOR result is her key bit string. Quantum secret sharing of an image of a mountain with a size of 79,800 bytes has been implemented, as shown in Fig.~\ref{demonstration}b. Only Bob and Charlie cooperate to recover the correct image, while a single player cannot recover the image and only obtains the complete noise map.

For the conference key agreement task, Alice, Bob, and Charlie are all honest participants and should have the same keys. This requirement can be realized with information-theoretical security if Alice's XOR result is published and Charlie changes his key to the same as Bob's through XOR operation. An image of a lake with a size of 139,500 bytes is adopted to implement quantum group encryption, as shown in Fig.~\ref{demonstration}c. Any of the three parties can individually obtain the correct image in the group encryption session.

The cryptographic tasks feature high efficiency and information-theoretical security on our quantum secure network using the current quantum technology. We remark that the trusted relay node Alice is required only in the private communication between Bob and Charlie on our quantum secure network. The other three tasks, digital signatures, secret sharing, and conference key agreement, are not required since all nodes are the task participants. Note that combining quantum secure direct communication and classical cryptography, secure quantum network can be constructed without quantum repeater as proposed in Ref.~\cite{long2022evolutionary}.

\section{CONCLUSIONS}\label{sec5}

In conclusion, we successfully demonstrated a full-function quantum secure network that meets all information security objectives, namely, confidentiality, integrity, authenticity, and non-repudiation. Particularly, we theoretically propose and experimentally implement an OTUH-QDS protocol that shows a 100-million-fold signature efficiency improvement. As such, digital signatures, which are critical in internet-based digital processing systems, are now promoted to be information-theoretically secure and commercially applicable by OTP, OTUH, and secret sharing. Our framework requires few resources to sign an almost arbitrarily long document, outperforming all previous protocols not only in signing efficiency but also in security. Of course, the full-function quantum secure network can be implemented by more advanced technology, such as a future quantum internet. Its successful implementation by a practical quantum secure network under current technology lays a firm foundation for a quantum secure layer of the current internet. Such a quantum secure internet, enabling main secure cryptographic tasks simultaneously, paves the way for the quantum age of the digital economy.

\section{METHODS}
\bigskip
\noindent
\textbf{Pre-distribution stage}

The pre-distribution stage ensures that each participant has two key bit strings and meets the secret sharing relationship $X_{a}=X_{b}\oplus X_{c}$ and $Y_{a}=Y_{b}\oplus Y_{c}$, which can be realized using quantum communication protocols with information-theoretical security.

There are two quantum key distribution links if quantum key distribution protocols are being observed. The Bob--Alice (Charlie--Alice) link will generate the symmetric quantum keys, denoted as $S_{ba}^{b}=S_{ba}^{a}$ $\left(S_{ca}^{c}=S_{ca}^{a}\right)$, and even dishonest Charlie (Bob) has no knowledge about it. Alice implements an XOR operation to obtain her new quantum key $S^{a}=S_{ba}^{a}\oplus S_{ca}^{a}$. Therefore, since Alice has all the knowledge of Bob and Charlie's keys, she can only be the signer, while Bob and Charlie can be the receivers. We remark that the XOR operation of Alice generates asymmetry between Alice and Bob.

Alice, Bob, and Charlie can directly generate the perfect correlation quantum keys $S_{a}=S_{b}\oplus S_{c}$ if quantum secret sharing protocols are being observed. Traditional quantum secret sharing protocols require that the dealer Alice is honest and the player Bob or Charlie can be allowed to be dishonest. The dishonest Bob and Charlie do not know $S_{a}$ and they can be the receiver. Note that since Alice can obtain all information of $S_{b}$ and $S_{c}$ if she is dishonest in performing traditional quantum secret sharing, Alice cannot be a receiver if traditional quantum secret sharing protocols are being used. However, if one adopts measurement-device-independent quantum secret sharing~\cite{fu2015longdistance}, all three participants will not know any information about others' quantum keys; anyone of them can be a receiver or signer.

\bigskip
\noindent
\textbf{One-time universal$_2$ hash function}

A collection $H$ of hash functions $h$: $\mathbb{S}$$\rightarrow$$\mathbb{T}$ is said to be universal$_{2}$~\cite{carter1979universal} if for every two different $x, y\in\mathbb{S}$, we have
\begin{equation}
	{\rm Pr}_{h\in H}[h(x)=h(y)]\le \frac{1}{|\mathbb{T}|}.
\end{equation}
This means that the universal$_2$ hash function can uniformly map the long documents to short hash values with a small collision probability. The random matrices belong to the universal$_2$ hash functions, which require $mn$ random bits for specifying hash functions (seen as an $mn$ Boolean matrix) to transform the $m$-bit document into an $n$-bit hash value. To reduce the cost of random bits, the Toeplitz matrix~\cite{menezes2018handbook}, which requires only $m+n-1$ random bits, is widely used in randomness extraction and privacy amplification, and its collision probability is $1/2^{n}$. Nevertheless, it still requires the length of random input bits to be longer than that of the document. Fortunately, the LFSR-based Toeplitz matrix~\cite{krawczyk1994lfsr} is the almost universal$_2$ hash function, where the hash function is determined by an irreducible polynomial $p(x)$ of degree $n$ over the Galois field ${\rm GF}(2)$ and $n$-bit random initial vector. The collision probability of the LFSR-based Toeplitz matrix~\cite{krawczyk1994lfsr} is $m/2^{n-1}$ (see Supplementary data). The initial vector and irreducible polynomial of the LFSR-based Toeplitz matrix are randomly changed for each signature, which is an important and novel requirement of our OTUH-QDS protocol.

\acknowledgments
We gratefully acknowledge the support from the Natural Science Foundation of Jiangsu Province (No. BK20211145), the Fundamental Research Funds for the Central Universities (No. 020414380182), the Key Research and Development Program of Nanjing Jiangbei New Aera (No. ZDYD20210101), and the Program for Innovative Talents and Entrepreneurs in Jiangsu (No. JSSCRC2021484).

\section{Supplementary data}

\bigskip\noindent\textbf{1. Cryptography toolbox}\\Here, we introduce threats faced by information processing (as shown in Fig.~\ref{simuandexp}), the corresponding information security objectives and classical cryptographic techniques to tackle such threats~\cite{schneier2015secrets}.
First, eavesdropping, which threatens the confidentiality of information, can be prevented using symmetric cryptography and asymmetric cryptography (i.e., public-key cryptography).
Second, tampering, which destroys the integrity of data, can be approached with a one-way hash function, message authentication code and digital signatures.
Third, disguise, in which the attacker pretends to be the real information sender, can deal with message authentication codes and digital signatures.
Finally, one may repudiate his or her certain behavior, and the digital signatures provide the efficacy of non-repudiation.

\begin{figure}[t]
	\centering
	\includegraphics[width=8.5cm]{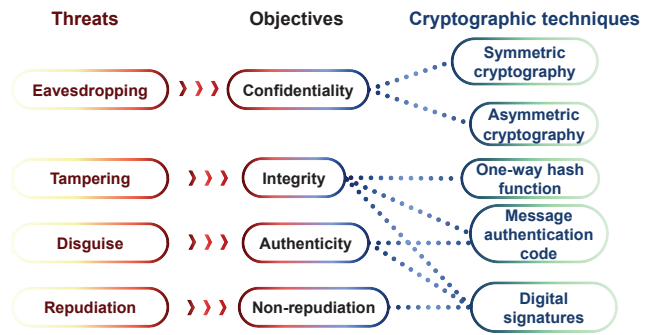}
	\caption{\textbf{Cryptography toolbox.}
	}
	\label{simuandexp}
\end{figure}

\bigskip
\noindent
\textbf{2. One-time pad}\\
Define that encryption algorithm $E$ maps plaintexts $m \in M$ to ciphertexts $c \in C$. According to Shannon's information theory, an algorithm $E$ is perfectly secret if $C$ and $M$ are independent, i.e., ${\rm Pr}(m, c) = {\rm Pr}(m)\times{\rm Pr}(c)$. The one-time pad~\cite{shannon1949communication} encryption satisfies the above definition, where a plaintext is paired using the XOR operation with a random secret key of the same length. Learning the ciphertext does not increase any plaintext information. In addition, each encryption requires a new and independent random key; hence, knowing details about the previous key does not help the attacker.

In our one-time universal$_2$ hashing quantum digital signatures (OTUH-QDS) scheme, we utilize the one-time pad to encrypt the hash value and the irreducible polynomial to acquire the signature so that we can completely conceal the information of (almost) universal$_2$ hash functions.

\begin{figure*}[t]
	\centering
	\includegraphics[width=16cm]{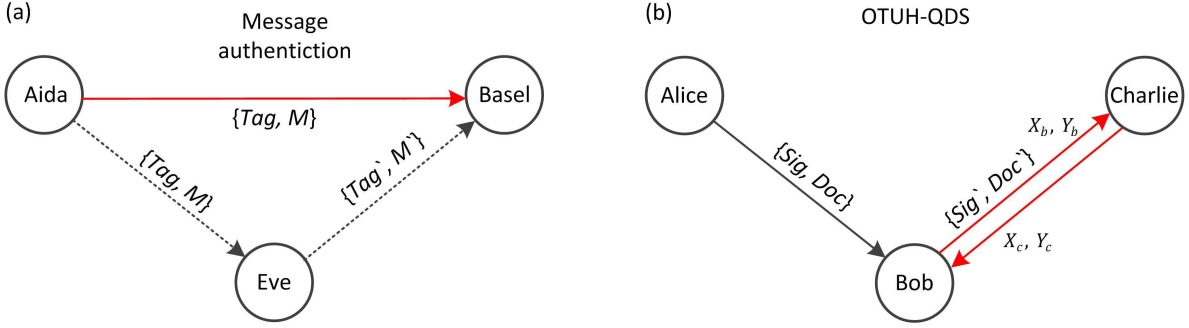}
	\caption{\textbf{Comparing  message authentication and digital signatures.} (a) Information-theoretically secure message authentication. Aida and Basel are the message sender and recipient, respectively, while Eve is the attacker. Eve tries to make Basel accept the tampered tag and message $\{Tag',M'\}$. The black dotted line represents Eve's attack process. (b) OTUH-QDS. For Bob's forgery attack, he tries to make Charlie accept the forged siganture and document $\{Sig',Doc'\}$. The red solid line represents the authentication classical channel, which has information-theoretical security by performing message authentication.
	}
	\label{f2}
\end{figure*}

\bigskip
\noindent
\textbf{3. LFSR-based Toeplitz hashing and message authentication}\\
The LFSR-based Toeplitz matrix~\cite{krawczyk1994lfsr} is the almost universal$_2$ hash function, where the hash function is determined by an irreducible polynomial $p(x)=x_{n}+a_{n-1}x^{n-1}+...+a_{1}x+a_{0}$ of degree $n$ over the Galois field ${\rm GF}(2)$ and $n$-bit random initial vector, just requiring a total of $2n$ bits. The randomness of the initial vector and irreducible polynomial together guarantees the security of LFSR-based Toeplitz hashing. The collision probability of LFSR-based Toeplitz hashing is bounded by $\epsilon=m/2^{n-1}$, which quantifies the upper bound on the probability of finding any two different documents with the same hash value. The LFSR-based Toeplitz hashing operation can be written as $h_{p,s}(M)$= $H_{nm} \cdot M=Hash$,  where  $p=(a_{n-1},a_{n-2},...,a_{1},a_{0})$ represents an irreducible polynomial and $s=(b_{n},b_{n-1},...,b_2,b_1)^T$ is the initial vector. $p$ and $s$ are random and determine the Toeplitz matrix $H_{nm}$ with $n$ rows and $m$ columns. $M=(M_{1},M_{2}...M_{m})^{T}$ is the message in the form of an $m$-bit vector.
The Toeplitz matrix can be written as $H_{nm}=(s,s_1,...,s_{m-1})$, where the initial vector $s=(b_n,b_{n-1},...,b_2,b_1)^T$ is the first column.
The LFSR, determined by $p$, will be performed $m-1$ times to generate $s_1,~s_2,~...,~s_{m-2}$ and $s_{m-1}$. To be specific, LFSR will shift down every element in the previous column,
and add a new element to the top of the column. For example, LFSR will first transform $s$ into $s_1=(b_{n+1},b_n,...,b_3,b_2)^T$, where $b_{n+1}=p \cdot s$, and likewise, transform $s_1$ to $s_2$. Then the $m$ vectors $s,s_1,...,s_{m-1}$ will together construct the Toeplitz matrix $H_{nm}=(s,s_1,...,s_{m-1})$, and the hash value of the message is $H_{nm} \cdot M=Hash$.

\emph{Failure probability of authentication.}---Here, we first provide the information-theoretical security proof of message authentication~\cite{krawczyk1994lfsr}. We remark that the security proof in the original literature is omitted~\cite{krawczyk1994lfsr}, we provide a detailed proof here, and the conclusions are consistent.
Suppose the message authentication scenario that an attacker Eve captures a tag and message $\{Tag,~M\}$ from the sender Aida, as shown in Fig.~\ref{f2}a. Note that the tag is acquired by using random secret key $Key$ to encrypt hash value $Hash$ with a one-time pad, i.e., $Tag=h_{p,s}(M)\oplus Key$. Eve cannot obtain any information of the Toeplitz hash function and the hash value due to the one-time pad. Eve can only randomly guess the initial vector $s$ and irreducible polynomial $p$. Eve can tamper with the message successfully if and only if he guesses a combination $\{t,~m\}$ with a very small probability that meets the relationship $h_{p,s}(m) = t$. Eve sends a new (tampered) tag and message $\{Tag'=Tag\oplus t,~M'=M\oplus m\}$ to the recipient Basel. In this case, the recipient will accept the message because of the relationship $h_{p,s}(M\oplus m)= h_{p,s}(M)\oplus h_{p,s}(m)= Tag\oplus t\oplus Key$.
It must be mentioned that $m\ne 0$ due to the requirement of a valid forge.
Thus, the failure probability $\epsilon_{\rm aut}$ of authentication by using LFSR-based Toeplitz hashing can be defined by the probability of one successfully choosing a combination $\{t,~m\}$ with the relationship $h(m) = t$ and $m\ne 0$,
\begin{equation}
	\epsilon_{\rm aut}={\rm Pr}[h_{p,s}(m) = t,~m\ne 0].
\end{equation}
We remark that the attacker Eve does not have any prior information about $p$ and $s$ and can only guess randomly. Moreover, before Eve sends the tag and message to the recipient, Eve also cannot be sure if he guessed correctly even if he exhausts all the possibilities, which is with the information-theoretical security~\cite{krawczyk1994lfsr}.
To quantify the failure probability $\epsilon_{\rm aut}$, we need to consider two cases, $t=0$ and $t \ne 0$. Note that $h_{p,s}(M)$ can be rewritten as
\begin{equation}
	h_{p,s}(M)=h_{p,M}(s)=  \sum^{m-1}_{i=0} M_{i+1}W^i\cdot s,
\end{equation}
where $W$ is the $n\times n$ circulant matrix with the first row being $p$. Denote the characteristic value of $W$ as $\lambda _i$ ($i=1,2,3,...,n$). One can diagonalize the matrix $\sum^{m-1}_{i=0} M_{i+1}W^i$, and the diagonal elements are $m(\lambda_i)$. After calculation, it can be verified that $p(x)$ is just the characteristic polynomial of $W$, which means that there is the relationship $p(\lambda_i)=0$.
For the first case $h_{p,M}(s)=t=0$, since $s\ne 0$,
$h_{p,M}(s)=0$ only if there exists a zero diagonal element in the diagonalized matrix, i.e., it has a zero characteristic value. Thus, there exists a $\lambda_i$ satisfying $m(\lambda_i)=0$.
This is equivalent to $p(x) | m(x)$ since $p(\lambda_i )=0$. The probability of such an event is at most the number of possible irreducible factors of $M(x)$ divided by the total number of irreducible polynomials of degree $n$. The former is at most $m/n$, and the latter is at least $2^{n-1}/n$~\cite{menezes2018handbook}.
Thus, the probability of $h_{p,s}(m)=t=0$ is at most $(m/n)/(2^{n-1}/n)=m/2^{n-1}$.
We include all the cases in which there exists a $\lambda_i$ satisfying $m(\lambda_i)=0$ in the first case, which is equivalent to the fact that the rank of $\sum^{m-1}_{i=0} M_{i+1}W^i $ is less than $n$. Thus, for the second case $h_{p,M}(s)=t\ne 0$, the rank of $\sum^{m-1}_{i=0} M_{i+1}W^i$ is equal to $n$. This means that $h_{p,M}(s)= \sum^{m-1}_{i=0} M_{i+1}W^i \cdot s$ is a bijection, i.e., one-to-one mapping. There are $2^n-1$ possible values of $s\ne 0$ corresponding to the $2^n-1$ different tag values, so the probability of $h_{p,s}(m)= t\ne 0$ is $1/(2^n-1)$.
Therefore, the upper bound on the failure probability of message authentication using LFSR-based Toeplitz hashing is~\cite{krawczyk1994lfsr}
\begin{equation}
	\epsilon_{\rm aut}=\max\left\{\frac{1}{2^n-1},\frac{m}{2^{n-1}}\right\}=\frac{m}{2^{n-1}}.
\end{equation}

We remark that message authentication is the premise and foundation for realizing the information-theoretical security of quantum key distribution~\cite{fung2010practical}, where the basis sift, error verification, and privacy amplification steps all require message authentication to ensure the information-theoretical security.

\bigskip
\noindent
\textbf{4. Secure OTUH-QDS against Bob's forgery attack}\\
Now, we will show that Bob's forgery attack in our OTUH-QDS protocol will be related to Eve's attack in secure message authentication, as shown in Fig.~\ref{f2}b. These two types of attacks are different but have many correlations. In message authentication, the message sender Adia and recipient Basel are honest and trust each other all the time. They will get together to against Eve's attack. A successful attack in message authentication is when Eve changes the tag and message and the recipient will accept it. In our OTUH-QDS, there are two types of Bob's forgery attacks since the signer Alice cannot be regarded as always honest. The first type is that Bob can generate a new signature and document if Alice has not signed a document at all. The second type is that Bob can change the signature and document if Alice has signed a document. Note that Bob does not have any information about the initial vector and the irreducible polynomial before he forwards the signed signature and document to charlie, which is the same as Eve in message authentication. We assume that the information-theoretically secure message authentication and OTUH-QDS will be performed many rounds to transfer multiple messages and sign multiple documents, respectively. Obviously, for the first round, the initial vector and irreducible polynomial are new and random. The failure probability is the same between Eve's attack in message authentication and Bob's forgery attack with the second type in OTUH-QDS since the attacker's purpose and the conditions for success are exactly the same, i.e., the failure probability are both $m/2^{n-1}$.

However, there is an important difference between message authentication and OTUH-QDS in the second round and beyond. In message authentication (Adia and Basel are always honest), the input initial vector and the irreducible polynomial can be fixed~\cite{chen2021integrated,fung2010practical} in later rounds since the attacker Eve cannot know the initial vector and the irreducible polynomial after message authentication has been performed. Bob also does not have any information about the initial vector and the irreducible polynomial before forwarding the signed document and signature to charlie. However, Charlie will forward his key bit strings to Bob after Charlie receives the document and signature forwarded by Bob. Thus, Bob can obtain all the information of the initial vector and the irreducible polynomial after the implementation of each round of digital signatures. To forbid Bob to exploit the information from the previous round to implement the attack. An important observation in our OTUH-QDS protocol is that the initial vector and the irreducible polynomial cannot be fixed and must be randomly updated in every round. In other words, the universal$_2$ hash function will only be used once and then be updated. Thus, we denote it as one-time universal$_2$ hashing, which is similar to the one-time pad.

\bigskip\noindent\textbf{5. Algorithms for testing irreducibility of polynomials over ${\rm GF}(2)$}\\
\emph{Irreducible polynomial.}---A crucial point is that in every round, the signer must randomly choose an irreducible polynomial in our OTUH-QDS scheme. In our protocol, Alice finishes this task with an $n$-bit quantum random number. Denote the format of the irreducible polynomial as $p(x)=x^n+a_{n-1} x^{n-1}+\cdots+a_1 x+a_0$, where $a_i=0$ or 1.
Then $p(x)$ can be characterized by an $n$-bit string $p=(a_{n-1},a_{n-2},\cdots,a_1,a_0)$, i.e., every bit of the string determines a corresponding coefficient of $p(x)$. To generate the irreducible polynomial, Alice first uses the $n$-bit random number to generate a polynomial $p_1 (x)$, and checks out whether $p_1 (x)$ is irreducible. If $p_1 (x)$ is irreducible, it will be used to generate the LFSR-based Toeplitz matrix. Otherwise, Alice utilizes a new $n$-bit random number to obtain a new string and generates a new polynomial and then examines whether it is an irreducible polynomial. Alice will repeat this step until she generates an irreducible polynomial. For example, choose $n=128$. Assume Alice first generates $p_1 (x)=x^{128}+x^{29}+x^{25}+x+1$. Then, she will find it reducible and successively generate and examine $p_2 (x)=x^{128}+x^{50}+x^{27}+x^2+1$ and $p_3 (x)=x^{128}+x^{29}+x^{27}+x^2+1$. Finally, she finds $p_3 (x)$ irreducible and chooses $p_3 (x)$ as the irreducible polynomial to generate the Toeplitz matrix. Note that for an irreducible polynomial $p(x)$, one always has $a_0=1$. Thus, we can use $n-1$ random bits $(a_{n-1},a_{n-2},\cdots,a_1,1)$ to generate the polynomial. We remark that the quantum random number used for generating the irreducible polynomial is produced locally by Alice's quantum random number generator.

\emph{The test algorithm.}---The following is the concrete method we employed to check whether a polynomial over ${\rm GF}(2)$ is irreducible or not. Suppose $p(x)$ is a polynomial of order $n$ in ${\rm GF}(2)$. According to~\cite{rabin1980probabilistic}, the necessary and sufficient condition for $p(x)$ being irreducible is:
\begin{equation}
	\left\{
	\begin{gathered}
		(A)~x^{2^n} \equiv x \mod p(x)
		\\[3pt]
		(B)~{\rm gcd}\left(x^{2^{\frac{n}{d}}}-x,~p(x)\right)=1,
	\end{gathered}
	\right.
\end{equation}
where $d$ is any prime factor of $n$ and ${\rm gcd}\left(f(x),~g(x)\right)$ represents the greatest common divisor (GCD) of $f(x)$ and $g(x)$. In our scheme, $n=128=2^7$,
i.e., $n$ only has one prime factor ``2". Thus, to verify condition (2), we just need to examine ${\rm gcd}\left(x^{2^{64}}-x,p(x)\right)=1$. To speed up our calculation, we utilize fast modular composition (FMC) algorithms and an extended Euclidean algorithm~\cite{von2013modern}. The FMC algorithms can calculate $x^{2^{128}}$ and $x^{2^{64}}$ mod $p(x)$ with high speed by calculating $x^{2^{2^7}}$ and $x^{2^{2^6}}$, while the extended Euclidean algorithm can quickly finish the GCD calculation. Thus, conditions (A) and (B) can be verified efficiently. In our simulation, we search 1000 irreducible polynomials, and on average, it takes 73.6 tests to find an irreducible polynomial,
consistent with the conclusion~\cite{krawczyk1994lfsr} that the total number of irreducible polynomials of order $n$ in ${\rm GF}(2)$ is at least $2^{n-1}/n$ and at most $2^{n}/n$~\cite{menezes2018handbook}.
We implement our simulation in a desktop with an Intel i5-10400 CPU (with RAM of 8 GB), and on average, it takes approximately 0.36 seconds to generate an irreducible polynomial of order 128 from a 128-bit random input.

\bigskip\noindent\textbf{6. Numerical simulation of our QDS via quantum key distribution element}\\
Considering the case that we have two quantum key distribution links between Alice-Bob and Alice-Charlie, the two links have the same loss. Alice and Bob share one set of secret keys $S_1$, while Alice and Charlie share the other set of secret keys $S_2$ by implementing quantum key distribution (QKD) with the key rate $R_{\rm QKD}$. For the requirement of secret sharing, Alice, Bob and Charlie utilize $K_a$, $K_b$ and $K_c$ as their correlation keys, respectively, where $K_a=S_1\oplus S_2$, $K_b=S_1$ and $K_c=S_2$. Therefore, the signature rate can be defined as $R_{\rm QDS} =R_{\rm QKD}/3n$, since one needs a $3n$-bit key to implement OTUH and OTP in our OTUH-QDS protocol.

\noindent\\
\emph{(6.1) Sending-or-not-sending twin-field quantum key distribution~\cite{wang2018TFlargemisalignment}.}\\
The secure key rate is given by
\begin{equation}
R_{\rm QKD}=2t(1-t)\mu e^{-\mu} Y_{1}\left[1-H\left(e_{1}^{\mathrm{ph}}\right)\right]-Q_ZfH(E^Z),
\end{equation}
where $t$ is the probability of sending the signal pulse at a signal window, $\mu$ is the intensity of the signal pulse, $Y_{1}$ is the counting rate of $Z_1$-windows, $Q_Z$ is the observed counting rate of $Z$-windows, $e_{1}^{\mathrm{ph}}=e_1^{X_1}$ is the phase-flip error rate for $Z_1$-windows, $E^Z$ is the bit error rate for $Z$-windows, and $f$ is the error correction efficiency.
The yield $Y_{1}$ and error rate $e_1^{X_1}$ can be estimated by exploiting the decoy-state method,
\begin{equation}
\begin{aligned}
Y_{1} \geq \frac{\mu / 2}{\mu\nu-\nu^{2}}\Big[&e^{\nu} (Q_{\nu0}+Q_{0\nu})-\frac{\nu^{2}}{\mu^{2}} e^{\mu} (Q_{\mu0}+Q_{0\mu})\\
&-2\frac{\mu^{2}-\nu^{2}}{\mu^{2}} Q_{00}\Big],
\end{aligned}
\end{equation}
\begin{equation}
e_1^{X_1} \leq \frac{1}{2 \nu Y_{1} }\left(e^{2 \nu} E_{\nu \nu}^{\mathrm{pm}} Q_{\nu \nu}^{\mathrm{pm}}-\frac{1}{2} Y_{0}\right),
\end{equation}
where $\nu$ is the intensity of the decoy pulse, $Q_{k_a,k_b}$ is the gain when Alice chooses intensity $k_a$ and Bob chooses intensity $k_b$ ($k_a,~k_b$ $\in\{\mu,~\nu,~0 \}$), $E_{\nu \nu}^{\mathrm{pm}}$ is the bit error rate, $Q_{\nu\nu}^{\mathrm{pm}}$ is the gain of intensity $\nu$ for both Alice and Bob after successful postselected phase-matching and $Y_{0}=Q_{00}$ is the counting rate of the vacuum state.
	
The gain $Q_{k_a,k_b}$ can be given as
\begin{equation}
\begin{aligned}
Q_{k_a,k_b}= 2(1-p_d)e^{-\frac{k_a+k_b}{2}\sqrt{\eta}}\Big[&I_0\left(\sqrt{k_ak_b}\sqrt{\eta}\right)\\
&-\left(1-p_d\right)e^{-\frac{k_a+k_b}{2}\sqrt{\eta}}\Big],
\end{aligned}
\end{equation}
where $\eta=\eta_d^{2}\times 10^{-\alpha L/10}$ is the total efficiency and  $\eta_d$, $\alpha$ and $L$ are the detector efficiency, attenuation coefficient and distance, respectively. $p_d$ is the dark count rate, and $I_0(x)$ is the modified Bessel function of the first kind.
The gain $Q_{\nu\nu}^{\mathrm{pm}}$ and bit error rate $E_{\nu \nu}^{\mathrm{pm}}$ can be given by
\begin{equation}
Q_{\nu \nu}^{\mathrm{pm}}=Q_{c, \nu \nu}^{\mathrm{pm}}+Q_{e, \nu \nu}^{\mathrm{pm}},
\end{equation}
\begin{equation}
E_{\nu \nu}^{\mathrm{pm}}=\left[e_{d}^{x} \mathrm{Q}_{c, \nu 	\nu}^{pm}+\left(1-e_{d}^{x}\right) Q_{e, \nu \nu}^{pm}\right] / \mathrm{Q}_{\nu \nu}^{pm},
\end{equation}
where $e_{d}^{x}$ is the misalignment rate of the X basis. Therein, the correct gain $Q_{c, \nu \nu}^{\mathrm{pm}}$ and incorrect gain $Q_{e, \nu \nu}^{\mathrm{pm}}$ are
\begin{equation}
\begin{aligned}
Q_{c, \nu \nu}^{\mathrm{pm}}  \approx &\frac{1-p_{d}}{\delta} \sqrt{\frac{\pi}{2 \nu \sqrt{\eta}}} \operatorname{erf}\left(\sqrt{\frac{\nu \sqrt{\eta}}{2}} \delta\right)\\
&-\left(1-p_{d}\right)^{2} e^{-2 \nu \sqrt{\eta}},
\end{aligned}
\end{equation}
and
\begin{equation}
\begin{aligned}
Q_{e, \nu \nu}^{\mathrm{pm}} \approx &\frac{1-p_{d}}{\delta} e^{-2 \nu \sqrt{\eta}} \sqrt{\frac{\pi}{2 \nu \sqrt{\eta}}} \operatorname{erfi}\left(\sqrt{\frac{\nu \sqrt{\eta}}{2}} \delta\right)\\
&-\left(1-p_{d}\right)^{2} e^{-2 \nu \sqrt{\eta}},
\end{aligned}
\end{equation}
where $\operatorname{erf}(x)$ and $\operatorname{erfi}(x)$ are the error function and imaginary error function, respectively.
The gain $Q_{Z}$ and bit error rate $E^Z$ can be given by
\begin{equation}
Q_Z=(1-t)^2Q_{00}+2t(1-t)Q_{0\mu}+t^2Q_{\mu\mu},
\end{equation}
and
\begin{equation}
E^Z=\frac{(1-t)^2Q_{00}+t^2Q_{\mu\mu}}{(1-t)^2Q_{00}+t(1-t)(Q_{0\mu}+Q_{\mu0})+t^2Q_{\mu\mu}}.
\end{equation}	
In the simulation, we set $p_d=10^{-8}$, $f=1.1$, $\eta_d=85 \%$, $e_d=2\%$ and $\alpha=0.167$ dB/km. The light intensities and probability $t$ are globally optimized.
	
\noindent\\
\emph{(6.2) Phase-matching quantum key distribution~\cite{ma2018phase}. 	}\\
Here we present the simulation details of the phase-matching quantum key distribution. The overall secure key rate is given by
\begin{equation}
R_{\rm QKD}= \frac{2}{D} Q_{\mu}[1-h(E^{\rm ph})-fh(E^{\rm b})],	
\end{equation}
where $Q_{\mu}$ is the total gain of the signal pulses, ${2}/{D}$ is the phase-sifting factor with $D=16$, $f$ is the error correction efficiency, $h(x)=-x\log_{2}{x}-(1-x)\log_{2}{(1-x)}$ is the binary entropy function, $E^{\rm ph}$ is the phase error rate and $E^{\rm b}$ is the bit error rate. The overall phase-error rate~\cite{zeng2020symmetry} is bounded by $E^{\rm ph}\le 1-q_1$, 	
where $q_1 = \mu e^{-\mu}Y_1/Q_{\mu}$. By using the decoy state method with three intensities, the yield of the single-photon component can be given as
\begin{equation}
Y_{1}\ge \frac{\mu}{\mu\nu-\nu^2}\left({Q}_{\nu} e^\nu - {Q}_{\mu} e^\mu \frac{\nu^2}{\mu^2}-\frac{\mu^2 - \nu^2}{\mu^2} {Q}_{0}\right).
\end{equation}
For simulation, the gain of intensity $k$ ($k\in\{\mu,~\nu,~0\}$) is  $Q_k=1-(1-2p_d)e^{-k\eta}$. The bit error rate is $E^{\rm b} = [p_d+\eta\mu(e_\Delta+e_d)]\frac{e^{-\eta\mu}}{Q_{\mu}}$, where $p_d$ is the dark count rate, $\eta=\eta_d\times 10^{-\alpha L/20}$ is the channel transmittance, $e_d$ is the extra misalignment error and the intrinsic error $e_\Delta= \sin^2 (\frac{\pi}{D}) $.
In the simulation, we set $p_d=10^{-8}$, $f=1.1$, $\eta_d=85\% $, $\alpha = 0.167$ dB/km  and $e_d=2\%$. The intensities $\mu$, $\nu$ are globally optimized.

\noindent\\
\emph{(6.3) Discrete-modulated continuous-variable quantum key distribution~\cite{lin2019asymptotic}.}\\
In the case of reverse reconciliation, the secret key rate under collective attacks in the asymptotic limit is given by $R_{\rm QKD}= p_{pass}[I(X;Z)-\max_{\rho\in\textbf{S}}\chi(Z:E)]$, where $I(X;Z)$ is the classical mutual information between Alice's string $X$ and the raw key string $Z$, $\chi(Z:E)$ is the information of Eve's knowledge about the raw key string $Z$ and $p_{pass}=1-\frac{1}{2}\int_{-\Delta_c}^{\Delta_c}P(q|0)dq-\frac{1}{2}\int_{-\Delta_c}^{\Delta_c}P(q|1)dq$ is the sifting probability.
$P(q|0)$ and $P(q|1)$ are probability distributions of Bob's homodyne detection conditioned that Alice generates bit 0 or 1.
The set $\textbf{S}$ contains all density operators compatible with experimental observations.
According to~\cite{lin2019asymptotic}, this key rate can be reformulated as a convex optimization problem, and the secret key rate is $R = \min_{\rho\in\textbf{S}}D(\mathcal{G}(\rho)||\mathcal{Z}[\mathcal{G}(\rho)])-p_{pass}\delta_{EC}$, where $D(\rho||\sigma)=\text{Tr}(\rho\log_2\rho)-\text{Tr}(\rho\log_2\sigma)$ is the quantum relative entropy.
$\mathcal{G}$ is a map describing classical postprocessing.
For the reverse reconciliation, we have $\mathcal{G}(\sigma) = K\sigma K^{\dagger}$, where $K=\sum_{z=0}^{1}\ket{z}_R\otimes(\ket{0}\bra{0}+\ket{1}\bra{1})_A\otimes(\sqrt{I_z})_B$ with $I_0=\int_{\Delta_c}^{\infty}dq\ket{q}\bra{q}$ and $I_1=\int^{-\Delta_c}_{-\infty}dq\ket{q}\bra{q}$.
We set $\Delta_c=0$ in our simulation.
$\mathcal{Z}$ is a pinching quantum channel, and we have $\mathcal{Z}(\rho)=\sum_jZ_j\rho Z_j$, where $Z_0=\ket{0}\bra{0}_R\otimes\mathcal{I}_{AB}$ and $Z_1=\ket{1}\bra{1}_R\otimes\mathcal{I}_{AB}$.
$\delta_{EC}=(1-\beta)H(Z)+\beta h(e)$ is the amount of information leakage per signal, where $e=\frac{1}{p_{pass}}\left( \frac{1}{2}\int^{-\Delta_c}_{\infty}P(q|0)dq+\frac{1}{2}\int^{\infty}_{\Delta_c}P(q|1)dq\right)$.
The convex optimization problem can be described as follows:
\begin{equation}
\begin{aligned}
minim&ize \quad D(\mathcal{G}(\rho)||\mathcal{Z}[\mathcal{G}(\rho)])\\
subje&ct \quad to\\
&Tr[\rho(\ket{i}\bra{i}_A \otimes \hat{q})] = p_x\langle\hat{q}\rangle_i,\\
&Tr[\rho(\ket{i}\bra{i}_A \otimes \hat{p})] = p_x\langle\hat{p}\rangle_i,\\
&Tr[\rho(\ket{i}\bra{i}_A \otimes \hat{n})] = p_x\langle\hat{n}\rangle_i,\\
&Tr[\rho(\ket{i}\bra{i}_A \otimes \hat{d})] = p_x\langle\hat{d}\rangle_i,\\
&Tr[\rho] = 1,\\
&\rho\geqslant 0,\\
&Tr_B[\rho] = \sum^{3}_{i,j=0}\sqrt{p_i p_j}\langle\phi_j|\phi_i\rangle\ket{i}\bra{j}_A,\\
\end{aligned}
\label{convexproblem}
\end{equation}
where $\hat{n}=\frac{1}{2}(\hat{q}^2+\hat{p}^2-1)=\hat{a}^{\dagger}\hat{a}$ and $\hat{d}=\hat{q}^2-\hat{p}^2=\hat{a}^2+(\hat{a}^{\dagger})^2$.
We consider a quantum channel with transmittance $\eta=10^{-0.167L/10}$ where $L$ is the distance and excess noise $\xi=0.01$.
In such a model, we have $P(q|0)=\frac{1}{\sqrt{\pi (1+\eta\xi)}}e^{-\frac{[\alpha-\sqrt{2\eta}q]^2}{1+\eta\xi}}$ and $P(q|1)=\frac{1}{\sqrt{\pi (1+\eta\xi)}}e^{-\frac{[\alpha+\sqrt{2\eta}q]^2}{1+\eta\xi}}$.
The expectation values can be given by
\begin{equation}
\begin{aligned}
\langle \hat{q} \rangle&= \sqrt{2\eta}Re(\alpha),\\
\langle \hat{p} \rangle&= \sqrt{2\eta}Im(\alpha),\\
\langle \hat{n} \rangle&= \eta|\alpha|^2 + \frac{\eta\xi}{2},\\
\langle \hat{d} \rangle&= \eta[\alpha^2 +(\alpha^*)^2].
\end{aligned}
\label{constraint}
\end{equation}
$\alpha$ is the amplitude of the signal, and we optimize $\alpha$ in the interval of $[0.35,0.6]$ with a step of 0.01.

\noindent\\
\emph{(6.4) Gaussian-modulated continuous-variable quantum key distribution~\cite{Lodewyck2007Quantum}.}\\
From an information-theoretic point of view, in the case of reverse reconciliation, Alice and Bob can distill perfectly correlated secret key bits provided that the amount of information they share $I_{AB}$ is higher than the information acquired by Eve $\chi_{BE}$ under collective attacks.
Therefore, the secret key rate is defined as $R = I_{AB}-\chi_{AB}$.
Considering the inefficiency of error correction, a factor $\beta$ is introduced as the reconciliation efficiency, and the secret key rate formula is $R_{\rm QKD} = \beta I_{AB}-\chi_{AB}$.
In our numerical simulation, we use $\beta=0.95$.
We calculate the mutual information using
\begin{equation}
I_{AB}=\frac{1}{2}\log_2\frac{V+\chi_{\rm tot}}{1+\chi_{\rm tot}},
\end{equation}
where $V$ is the variance of the thermal state observed at Alice's lab, and $\chi_{tot}$ is the total excess noise added between Alice and Bob.
Here, we detail the formula of $\chi_{\rm tot}$ as follows:
\begin{equation}
\chi_{tot}=\frac{1}{\eta T(1+v_{el})+\xi-1},
\end{equation}
where $T$ is the transmission of the quantum channel, $\xi$ is the excess noise, $v_{\rm el}$ is the electronic noise and $\eta$ is the detection losses.
To draw the line, we set $\eta=0.85$, $v_{\rm el} = 0$, $\xi=0.01$, $T = 10^{-0.167L/10}$ and $V$ is optimized.
$L$ is the distance.
Under collective attacks, Eve's accessible information is upper bounded by the Holevo quantity $\chi_{BE}$ satisfying
\begin{equation}
\begin{aligned}
\chi_{BE}=&G\left(\frac{\lambda_1-1}{2}\right)+G\left(\frac{\lambda_2-1}{2}\right)-G\left(\frac{\lambda_3-1}{2}\right)\\
&-G\left(\frac{\lambda_4-1}{2}\right),
\end{aligned}
\end{equation}
where $G(x)=(x+1)\text{log}_2(x+1)-x\log_2 x$, and $\{\lambda_i\}$ are the symplectic eigenvalues of the covariance matrix characterizing the quantum state.
Calculation shows that $\lambda_{1,2}^2=\frac{1}{2}[A\pm\sqrt{A^2-4B}]$, where $A = V^2(1-2T)+2T+T^2(V+\xi-1+\frac{1}{T})^2$ and $B=T^2(V/T-V+V\xi+1)^2$ and $\lambda_{1,2}^2=\frac{1}{2}[C\pm\sqrt{C^2-4D}]$, where $C=\frac{V\sqrt{B}+T(V+1/T-1+\xi)+A/\eta(1+v_{\rm el})-A}{T(V+\chi_{\rm tot})}$ and $D = \sqrt{B}\frac{V+\sqrt{B}[1/\eta(1+v_{\rm el})-1]}{T(V+\chi_{\rm tot})}$.
	
\noindent\\
\emph{(6.5) Measurement-device-independent quantum key distribution~\cite{lo2012measurement}.	}\\
For the measurement-device-independent quantum key distribution, we consider Alice and Bob sending a coherent state and the ideal case where only the channel loss and detector efficiency are taken into consideration. The secret key rate is given by
\begin{equation}
R_{\rm QKD}=\frac{1}{2e^2}\eta^2,
\end{equation}
where $e$ is the natural constant, $\eta=\eta_d\times 10^{-\alpha L/20}$ is the total channel transmittance and $L$ is the distance between Alice and Bob. We set $\eta_d= 85 \% $, $\alpha = 0.167$ dB/km.

\bigskip
\noindent\textbf{7. Numerical simulation of our QDS via quantum secret sharing element}\\
\noindent\\
Different from quantum key distribution, quantum secret sharing can directly offer secret sharing correlation with unconditional security via quantum laws. In quantum secret sharing of three users, Alice is the dealer, while both Bob and Charlie are players. Even though one of Bob and Charlie is dishonest, they can also generate the secure correlation key $K_{a}=K_{b}\oplus K_{c}$. We assume that the distances of Alice-Bob and Alice-Charlie are the same. The signature rate can be defined as $R_{\rm QDS}=R_{\rm QSS}/3n$, since one needs a $3n$-bit key for OTUH and OTP.

\noindent\\
\emph{7.1) Prepare-and-measure quantum secret sharing~\cite{hillery1999quantum}.}\\
Using the Greenberger-Horne-Zeilinger (GHZ) state can realize quantum secret sharing directly. Here, we consider an equivalent prepare-and-measure protocol.
In our simulation, we only consider the photon loss in the quantum channel and detector. Therefore, the secret key rate is given by
\begin{equation}
R_{\rm QSS} = \eta^{2},
\end{equation}
where $\eta=\eta_d \times10^{-\alpha L/10}$, $\eta_d=85 \%$ and $\alpha = 0.167$ dB/km. $L$ is the distance between Alice and Bob (Charlie).

\noindent\\
\emph{(7.2) Measurement-device-independent quantum secret sharing~\cite{fu2015longdistance}.}\\
By exploiting the postselected GHZ state, one can also realize quantum secret sharing. We only consider the photon loss from quantum channels and detectors here. The secret key rate of using a single-photon source can be given by
\begin{equation}
R_{\rm QSS} = \frac{1}{4}\eta_d\eta^{2}.
\end{equation}
Here, we have $\eta=\eta_d\times 10^{-\alpha L/10}$, $\eta_d=85 \%$, $\alpha = 0.167$ dB/km and assume that there is no loss of Alice's photon. The factor $ 1/4$ comes from only two of eight GHZ states that can be identified. $L$ is the distance between Alice and Bob (Charlie).

\noindent\\
\emph{(7.3) Round-robin quantum secret sharing~\cite{gu2021secure}.}\\	
The secret key rate per pulse of the round-robin quantum secret sharing with a twin-field for sending $d$ coherent pulses each train in the case of the inside adversary can be written as	
\begin{equation}\label{eq9}
\begin{aligned}
R_{\rm QSS}=\frac{1}{d}\{\hat{Q}[1-h(\hat{e}_{\rm p})]-Qfh(e_{\rm b})\},
\end{aligned}
\end{equation}
where we have $\hat{Q}=\min\{Q_{A},Q_{B}\}$. $Q_A$ and $Q_B$ are the gains of Charlie's successful detection from Alice and Bob, respectively. The phase error rate is
\begin{equation}\label{eq10}
\begin{aligned}
\hat{e}_{p}=\frac{e_{\rm src}}{\hat{Q}}+\left(1-\frac{e_{\rm src}}{\hat{Q}}\right)\frac{n_{\rm th}}{d-1}.
\end{aligned}
\end{equation}
The average gain $Q$ and bit error rate $e_{\rm b}$ of each train can be written as
\begin{equation}
\begin{aligned}
Q=\frac{1}{2}[1-(1-dp_{d})e^{-2\mu \eta}],
\end{aligned}
\end{equation}
and
\begin{equation}
\begin{aligned}
e_{\rm b}=\frac{e_{d}(1-e^{-2\mu {\eta}})+dp_{d}e^{-2\mu {\eta}}/2}{1-(1-dp_{d})e^{-2\mu {\eta}}},
\end{aligned}
\end{equation}
where ${\eta}=\eta_{d}\times 10^{-\alpha L/10}$ is the efficiency between Alice and Bob (Charlie). In addition, we have the gains $Q_{A}=Q_{B}=Q/2$ due to the symmetry. Let $n$ be the total photon number in a train of optical pulses with total intensity $\mu$. Then, the probability of finding more than $n_{\rm th}$ photons in a train of optical pulses can be written as
\begin{equation}
\begin{aligned}
{\rm Pr}(n>n_{\rm th})=e_{\rm src}:=1-\sum_{n=0}^{n_{\rm th}}\frac{e^{-\mu}\mu^{n}}{n!},
\end{aligned}
\end{equation}
where $n_{\rm th}$ is an integer constant chosen in this protocol. For the simulation, we set $p_d=10^{-8}$, $f=1.1$, $\eta_d=85\% $, $\alpha = 0.167$ dB/km and $e_d=2\%$. The intensity $\mu$, the selected integer $n_{\rm th} $ and the number of optical pulses $d$ are globally optimized.

\noindent\\
\emph{(7.4) Single-qubit quantum secret sharing~\cite{schmid2005experimental}.}\\		
For the single-qubit quantum secret sharing protocol with a single photon source, the final key rate can be easily derived based on phase-error correction
\begin{equation}
R_{\rm QSS}=Q_1[1-h(e_{\rm p})-fh(e_{\rm b})],
\end{equation}
where $e_{\rm p}=e_{\rm b}=p_d (1-2 e_d)/Q_1 + e_d$ and $Q_1 = \mu e^{-\mu} (2 p_{d}+\eta-2 p_{d} \eta)$. $\eta =\eta_d \times 10^{-\alpha \cdot 2L/10}$ is the total transmittance; $2L$ is the total distance. The intensity $\mu$ is globally optimized. We set $p_d=10^{-8}$, $f=1.1$, $\eta_d=85\% $, $\alpha = 0.167$ dB/km and $e_d=2\%$.

\begin{table*}[t]
\centering	
\caption{List of the experimental data used for one-time secure key generation.}
\setlength{\tabcolsep}{12mm}{
\begin{tabular}{c|c|c}
\hline
\hline
$~$ ~~~~~&  ~~~Alice-Bob (101 km)~~~&  ~~~Alice-Charlie (126 km)~~~  \\
\hline
~~~$n_{\mu}^z$ ~~~~~& 4616266~~~~~ &  5225387~~~~~  \\
~~~$n_{\nu}^z$ ~~~~~& 258671~~~~~ & 296285~~~~~  \\
~~~$n_{\omega}^z$ ~~~~~& 394425~~~~~ & 437496~~~~~  \\
~~~$n_0^z$ ~~~~~& 11189~~~~~ & 14080~~~~~  \\
~~~$n_{\mu}^x$ ~~~~~& 1035046~~~~~ & 1311484~~~~~  \\
~~~$n_{\nu}^x$ ~~~~~& 58107~~~~~ & 78890~~~~~  \\
~~~$n_{\omega}^x$ ~~~~~& 87317~~~~~ & 110362~~~~~  \\
~~~$n_{0}^x$ ~~~~~& 3126~~~~~ & 8168~~~~~  \\
~~~$m_{\omega}^z$ ~~~~~& 3446~~~~~ & 9401~~~~~  \\
\hline
\hline
\end{tabular}}
\label{Table1}
\end{table*}

\noindent\\
\emph{(7.5) Differential-phase-shift quantum secret sharing~\cite{gu2021differential}.}\\	
The final key rate for differential-phase-shift quantum secret sharing using a twin-field is bounded by
\begin{equation}
R_{\rm QSS} = Q_{\mu}[ -(1-2\mu)\log_2(P_{\rm co}) - fh(E_{\mu}) ],
\end{equation}
where $Q_{\mu}$ is the gain of the whole system. $h(x)=-x\log_{2}(x)-(1-x)\log_{2}(1-x)$ is the Shannon entropy, and $f$ is the error correction efficiency. $P_{\rm co}$ is the upper bound of collision probability when considering individual attacks, which can be concluded as $P_{\rm co} = 1-E_{\mu}^2-{(1-6E_{\mu})^2}/{2}$.
The total gain and the total error rate with an intensity of $\mu$ are given by	
\begin{equation}\label{qmuemu}
\begin{aligned}
&Q_\mu = 1 - (1-2p_d)e^{-\mu \eta},\\
&E_{\mu}Q_{\mu} = e_d Q_\mu + \left(\frac{1}{2}-e_d\right) 2p_d e^{-\mu \eta},\\
\end{aligned}
\end{equation}
where $e_d$ is the misalignment error rate of detectors.
$\eta=\eta_d \times 10^{-{\alpha L}/{10}}$ is the efficiency between Alice and Bob (Charlie), where $\eta_d$ is the detection efficiency of Charlie's detectors. The intensity is globally optimized. We set $p_d=10^{-8}$, $f=1.16$, $\eta_d=85\% $, $\alpha = 0.167$ dB/km and $e_d=2\%$.

\bigskip\noindent\textbf{8. Experimental details}\\In the transmitting end, a master laser generates a repetition rate of 200 MHz and phase-randomized laser pulses 1.6 ns-wide at 1550.12 nm. To avoid using a phase modulator, we utilize two slave lasers to generate relative phases 0 and $\pi$ by using the quantum properties of the beam splitter. An asymmetric interferometer with a 2 ns time delay divides each master pulse into two pairs of optical pulses with relative phases 0 and $\pi$. Then, the two pairs of optical pulses are
injected into two slave lasers through the optical circulator. With the help of controlling the trigger electrical signal of two slave lasers, one generates a quantum signal only in the first time-bin or the second time-bin to constitute the $Z$ basis, and one prepares a quantum signal both in two time-bins with a 0 or $\pi$ phase difference to constitute the $X$ basis. A 50 GHz bandwidth fiber Bragg grating is exploited to precompensate for pulse broadening and to remove extra spurious emission. The 2 ns-wide synchronization pulses with repetition rates of 100 kHz are transmitted via the quantum channel by using wavelength division multiplexed. The slave pulse width is 400 ps, which is much larger than the 10 ps timing resolution of the programmable delay chip, which means that the time consistency can be accurately calibrated. The spectral consistency is naturally satisfied through the laser seeding technique~\cite{comandar2016quantum}.

In the receiving end, a 30:70 biased beam splitter is used to perform passive basis detection after a wavelength division demultiplexer. A probability of $30\%$ is measured in phase interference and the probability of $70\%$ is used to receive in the time basis. A Faraday-Michelson interferometer realizes the phase measurement, where phase drift is compensated in real time by using the phase shifter. Two single photon detectors are used to measure the first time bin and the second time bin in the Z basis. Another two single photon detectors are used to measure the 0 and $\pi$ phase difference in the X basis. The total insertion losses of the time and phase bases are 4.25 and 8 dB, respectively.
The efficiency of single-photon detectors is $20\%$ at a 160 dark count per second. To decrease the after-pulse probability, we set the dead times to 10 and 25 $\mu$s for the Bob-Alice and Charlie-Alice links, respectively.

A four-intensity decoy-state protocol~\cite{Wang2005Beating,Lo:2005:Decoy,Yu:2016:Reexamination} is adopted, where the intensities of the $Z$ basis are set as $\mu=0.35$ and $\nu=0.15$, and the intensity of the $X$ basis is $\omega=0.3$. The intensity of the vacuum state is 0, which does not contain any basis information. The corresponding probabilities are $p_{\mu}=0.78$, $p_{\nu}=0.1$, $p_{\omega}=0.08$ and $p_{0}=0.04$. Thereinto, the amplitude modulator generates two different intensities, and the intensity of $\omega$ is double that of $\nu$ since it has two pulses in the $X$ basis.

The error correction and privacy amplification are carried out using a field-programmable gate array. Each time, privacy amplification will be performed after accumulating data approximately to the size of 4 Mb via approximately ten times of error correction, where the data size excludes the amount of information leaked in error correction. For the link of Alice-Bob with 101 km (Alice-Charlie with 126 km), one needs to accumulate approximately 153 (560) seconds of data for privacy amplification to extract a secure key. Here, we only list the experimental data of one set and calculation results related to privacy amplification, as shown in Tables~\ref{Table1} and~\ref{Table2}.

\begin{table*}[]
\centering	
\caption{List of the calculation results used for one-time secure key generation.}
\setlength{\tabcolsep}{10mm}{
\begin{tabular}{c|c|c}
\hline
\hline
$~$ ~~~~~&  ~~~Alice-Bob (101 km)~~~&  ~~~Alice-Charlie (126 km)~~~  \\
\hline
~~~$R_{\rm QKD}$ (bps) ~~~~~& 6021~~~~~ &  470~~~~~  \\
~~~$\Lambda$ (s) ~~~~~& 152.8~~~~~ & 561~~~~~  \\
~~~$\underline{s}_{0}^{zz}$ ~~~~~& 159429~~~~~ & 203152~~~~~  \\
~~~$\underline{s}_{1}^{zz}$ ~~~~~& 2756924~~~~~ & 3158726~~~~~  \\
~~~$\lambda_{\rm EC}$ ~~~~~& 1225992~~~~~ & 1659948~~~~~  \\
~~~$\ell$ ~~~~~& 920042~~~~~ & 263482~~~~~  \\
~~~$\overline{\phi}_{1}^{zz}$ ~~~~~& 4.83\%~~~~~ & 9.57\%~~~~~  \\
~~~$\varepsilon_{\rm \sec}$ ~~~~~& $10^{-10}$~~~~~ & $10^{-10}$~~~~~  \\
\hline
\hline
\end{tabular}}
\label{Table2}
\end{table*}

To compare with the experimental results, we use the relevant experimental parameters to simulate the secure key rate, as shown in Table III. The length of the final key, which is  $\varepsilon_{\rm cor}$-correct and $\varepsilon_{\rm sec}$-secret, can be given by~\cite{lim2014concise,yin2020experimental}
\begin{equation}\label{eq1}
\begin{split}
\ell = &\underline{s}_0^{zz}+\underline{s}_1^{zz}\left[1-h\left( \overline{\phi}_1^{zz} \right)\right]-\lambda_{\rm EC}-\log_2\frac{2}{\varepsilon_{\rm cor}}-6\log_2\frac{22}{\varepsilon_{\rm sec}},
\end{split}
\end{equation}
where~$h(x):=-x\log_2x-(1-x)\log_2(1-x)$, and $\overline{x}$ ($\underline{x}$) denotes the upper (lower) bound of the observed value $x$. Using the decoy-state method for finite sample sizes, the expected numbers of vacuum events $\underline{s}_0^{zz^*}$ and single-photon events $\underline{s}_1^{zz^{*}}$ can be written as
\begin{equation}
\begin{aligned}\label{eq2}
\underline{s}_0^{zz^*}\geq &( e^{-\mu}p_\mu+ e^{-\nu}p_\nu) \frac{\underline{n}_0^{z^*}}{p_0},\\
\end{aligned}
\end{equation}
and
\begin{equation}
\begin{aligned}\label{eq3}
\underline{s}_1^{zz^*}\geq&\frac{\mu^{2} e^{-\mu}p_\mu+\mu\nu e^{-\nu}p_\nu}{\mu\nu-\nu^2}\\		
&\times\left(e^\nu \frac{\underline{n}_\nu ^{z^*}}{p_\nu}-\frac{\nu^2}{\mu^2}e^\mu \frac{\overline{n}_\mu^{z^*}}{p_\mu}-\frac{\mu^2-\nu^2}{\mu^2}\frac{\overline{n}_0^{z^*}}{p_0}\right),
\end{aligned}
\end{equation}
where $n_k^{z(x)}$ is the count of $k$ ($k\in\{\mu,~\nu,~\omega\}$) intensity pulse measured in the $Z (X)$ basis, and $x^{*}$ is the corresponding expected value of given observed value $x$. The upper and lower bounds can be acquired~\cite{yin2020tight} by using the variant of the Chernoff bound, $\overline{x}^{*}=x+\beta+\sqrt{2\beta x+\beta^{2}}$ and
$\underline{x}^{*}=x-\frac{\beta}{2}-\sqrt{2\beta x+\frac{\beta^{2}}{4}}$, where  $\beta=\ln\frac{22}{\varepsilon_{\rm sec}}$.
The expected number of single-photon events $\underline{s}_1^{xx^{*}}$ in $\cX_{\omega}$ can be given by~\cite{Yu:2016:Reexamination}
\begin{equation}
\begin{aligned}\label{eq5}
\underline{s}_1^{xx^*}\geq&\frac{\mu\omega e^{-\omega}p_\omega}{\mu\nu-\nu^2}\left(e^\nu \frac{\underline{n}_\nu ^{x^*}}{p_\nu}-\frac{\nu^2}{\mu^2}e^\mu \frac{\overline{n}_\mu^{x^*}}{p_\mu}-\frac{\mu^2-\nu^2}{\mu^2}\frac{\overline{n}_0^{x^*}}{p_0}\right).\\
\end{aligned}
\end{equation}
In addition, the expected number of bit errors $\underline{t}_1^{xx^{*}}$ associated with the single-photon event in $\cX_{\omega}$ is $\overline{t}_1^{xx}\leq m_{\omega}^{x}-\underline{t}_{0}^{xx}$, with $\underline{t}_{0}^{xx^{*}}=\frac{e^{-\omega}p_{\omega}}{2p_{0}}\underline{n}_{0}^{x^*}$.
For a given expected value, the upper and lower bounds of the observed value can be given as $\overline{x}=x^{*}+\frac{\beta}{2}+\sqrt{2\beta x^{*}+\frac{\beta^{2}}{4}}$ and $\underline{x}=x^{*}-\sqrt{2\beta x^{*}}$, respectively.
\begin{table*}[]
\centering	
\caption{Simulation parameters of the link between Alice-Bob (Charlie). $e_{d}^{z}$ and $e_{d}^{x}$ are the misalignment rates of the $Z$ and $X$ bases, and $t_{\rm dead}$ is the dead time of the detector.}
{
\begin{tabular}{c|c|c|c|c|c|c|c}
\hline
\hline

~~~Clock frequency~~~ &~~~$\eta_d$ ~~~&  ~~~$p_d$& ~~~$e_{d}^{z}$~~~ & ~~~$e_{d}^{x}$~~~ &~~~ $t_{\rm dead}$~~~&~~~ $\varepsilon_{\rm cor}$ ~~~& ~~~ $f$   \\

\hline

~~~$2\times10^{8}$~~~ &	~~~$20 \%$ ~~~&~~~ $0.8\times10^{-6}$ ~~~&~~~ 2.3\% (2.2\%)~~~ & ~~~1.9\% (2.4\%) ~~~ &~~~10 (25) $\mu$s~~~ &~~~$10^{-15}$~~~&~~~ 1.42  \\

\hline
\hline
\end{tabular}}
\label{Table3}
\end{table*}
By using random sampling without replacement, the phase error rate in the $Z$ basis is
\begin{equation}\label{eq9}
\overline{\phi}_1^{zz}=\frac{\overline{t}_1^{xx}}{\underline{s}_1^{xx}}+\gamma^{U}\left(\underline{s}_1^{zz},\underline{s}_1^{xx},\frac{\overline{t}_1^{xx}}{\underline{s}_1^{xx}},\frac{\varepsilon_{\rm sec}}{22}\right),
\end{equation}
where we have $\gamma^{U}(n,k,\lambda,\epsilon)=\frac{\frac{(1-2\lambda)AG}{n+k}+\sqrt{\frac{A^2G^2}{(n+k)^2}+4\lambda(1-\lambda)G}}{2+2\frac{A^2G}{(n+k)^2}}$, with $A=\max\{n,k\}$ and $G=\frac{n+k}{nk}\ln{\frac{n+k}{2\pi nk\lambda(1-\lambda)\epsilon^{2}}}$.

\noindent\\
{\emph{Experimental demonstration the single-bit-type QDS of Ref.~\cite{Amiri:Secure:2016}.}}\\	
All the shared keys do not require error correction, and privacy amplification, i.e., $n_{\mu}$, can be used for signature. For a one-bit message, suppose that $2L$ bits of keys are used. Then the security level of the signature can be bounded by~\cite{an2019practical}
\begin{equation}
	\epsilon=max(P(honest ~abort),P(repudiation),P(forge)),
\end{equation}

\begin{equation}
	P(honest ~abort)=2e^{-(s_a-\overline{E})^2L},
\end{equation}

\begin{equation}
	P(repudiation)=2e^{-(s_v-s_a)^2L},
\end{equation}

\begin{equation}
	P(forge)=2e^{-(p_{e}-s_v)^2L}.
\end{equation}
where we choose $s_a=\overline{E}+\frac{p_e-\overline{E}}{3}, s_v=\overline{E}+\frac{2(p_e-\overline{E})}{3}$, and $\overline{E}$ is the upper bound bit error rate of the signal state. $p_e$ can be determined by
\begin{equation}
	\underline{c}^{zz}_{0}+\underline{c}^{zz}_{1}[1-h(\overline{\phi}^{zz}_{1})]=h(p_e)
\end{equation}
where $\underline{c}^{zz}_{i}=\underline{s}^{zz}_{i}/n_\mu$.

Note that it is different between the Bob-Alice link and Charlie-Alice link. The parameters $\overline{E}$ and $p_e$ should choose the maximal and minimum values, respectively. By using the experimental data, we can estimate $p_e=7.06\%$, $s_v=5.79\%$, $s_a=4.52\%$ and $\overline{E}=3.24\%$. Thus, to sign one bit, we need raw keys with $2L=1.09\times10^{6}$~($4.66\times 10^5$) bits, and the probability of honest abort, repudiation and forge are $P(honest~abort)=P(repudiation)=P(forge)=\hat{\epsilon}=10^{-38}~(10^{-16})$, respectively. When the length of the document is one megabit, the signature security bound $\hat{\epsilon}=10^{-16}$ is $\epsilon=1-(1-\hat{\epsilon}^{10^6})\approx 10^{6}\cdot \hat{\epsilon}=10^{-32}~(10^{-10})$.


\begin{thebibliography}{60}%
\makeatletter
\providecommand \@ifxundefined [1]{%
 \@ifx{#1\undefined}
}%
\providecommand \@ifnum [1]{%
 \ifnum #1\expandafter \@firstoftwo
 \else \expandafter \@secondoftwo
 \fi
}%
\providecommand \@ifx [1]{%
 \ifx #1\expandafter \@firstoftwo
 \else \expandafter \@secondoftwo
 \fi
}%
\providecommand \natexlab [1]{#1}%
\providecommand \enquote  [1]{``#1''}%
\providecommand \bibnamefont  [1]{#1}%
\providecommand \bibfnamefont [1]{#1}%
\providecommand \citenamefont [1]{#1}%
\providecommand \href@noop [0]{\@secondoftwo}%
\providecommand \href [0]{\begingroup \@sanitize@url \@href}%
\providecommand \@href[1]{\@@startlink{#1}\@@href}%
\providecommand \@@href[1]{\endgroup#1\@@endlink}%
\providecommand \@sanitize@url [0]{\catcode `\\12\catcode `\$12\catcode
  `\&12\catcode `\#12\catcode `\^12\catcode `\_12\catcode `\%12\relax}%
\providecommand \@@startlink[1]{}%
\providecommand \@@endlink[0]{}%
\providecommand \url  [0]{\begingroup\@sanitize@url \@url }%
\providecommand \@url [1]{\endgroup\@href {#1}{\urlprefix }}%
\providecommand \urlprefix  [0]{URL }%
\providecommand \Eprint [0]{\href }%
\providecommand \doibase [0]{https://doi.org/}%
\providecommand \selectlanguage [0]{\@gobble}%
\providecommand \bibinfo  [0]{\@secondoftwo}%
\providecommand \bibfield  [0]{\@secondoftwo}%
\providecommand \translation [1]{[#1]}%
\providecommand \BibitemOpen [0]{}%
\providecommand \bibitemStop [0]{}%
\providecommand \bibitemNoStop [0]{.\EOS\space}%
\providecommand \EOS [0]{\spacefactor3000\relax}%
\providecommand \BibitemShut  [1]{\csname bibitem#1\endcsname}%
\let\auto@bib@innerbib\@empty
\bibitem [{\citenamefont {Fedorov}\ \emph {et~al.}(2018)\citenamefont
  {Fedorov}, \citenamefont {Kiktenko},\ and\ \citenamefont
  {Lvovsky}}]{fedorov2018quantum}%
  \BibitemOpen
  \bibfield  {author} {\bibinfo {author} {\bibfnamefont {A.}~\bibnamefont
  {Fedorov}}, \bibinfo {author} {\bibfnamefont {E.}~\bibnamefont {Kiktenko}},\
  and\ \bibinfo {author} {\bibfnamefont {A.}~\bibnamefont {Lvovsky}},\
  }\bibfield  {title} {\bibinfo {title} {Quantum computers put blockchain
  security at risk},\ }\href@noop {} {\bibfield  {journal} {\bibinfo  {journal}
  {Nature}\ }\textbf {\bibinfo {volume} {563}},\ \bibinfo {pages} {465}
  (\bibinfo {year} {2018})}\BibitemShut {NoStop}%
\bibitem [{\citenamefont {Menezes}\ \emph {et~al.}(2018)\citenamefont
  {Menezes}, \citenamefont {Van~Oorschot},\ and\ \citenamefont
  {Vanstone}}]{menezes2018handbook}%
  \BibitemOpen
  \bibfield  {author} {\bibinfo {author} {\bibfnamefont {A.~J.}\ \bibnamefont
  {Menezes}}, \bibinfo {author} {\bibfnamefont {P.~C.}\ \bibnamefont
  {Van~Oorschot}},\ and\ \bibinfo {author} {\bibfnamefont {S.~A.}\ \bibnamefont
  {Vanstone}},\ }\href@noop {} {\bibinfo {title} {Handbook of applied
  cryptography}} (\bibinfo {year} {2018})\BibitemShut {NoStop}%
\bibitem [{\citenamefont {Wang}\ and\ \citenamefont
  {Yu}(2005)}]{wang2005break}%
  \BibitemOpen
  \bibfield  {author} {\bibinfo {author} {\bibfnamefont {X.}~\bibnamefont
  {Wang}}\ and\ \bibinfo {author} {\bibfnamefont {H.}~\bibnamefont {Yu}},\
  }\href@noop {} {\bibinfo {title} {How to break md5 and other hash functions}}
  (\bibinfo {year} {2005})\BibitemShut {NoStop}%
\bibitem [{\citenamefont {Wang}\ \emph {et~al.}(2005)\citenamefont {Wang},
  \citenamefont {Yin},\ and\ \citenamefont {Yu}}]{wang2005finding}%
  \BibitemOpen
  \bibfield  {author} {\bibinfo {author} {\bibfnamefont {X.}~\bibnamefont
  {Wang}}, \bibinfo {author} {\bibfnamefont {Y.~L.}\ \bibnamefont {Yin}},\ and\
  \bibinfo {author} {\bibfnamefont {H.}~\bibnamefont {Yu}},\ }\bibfield
  {title} {\bibinfo {title} {Finding collisions in the full sha-1},\ }in\
  \href@noop {} {\emph {\bibinfo {booktitle} {Advances in Cryptology - CRYPTO
  2005: 25th Annual International Cryptology Conference, Santa Barbara,
  California, USA, August 14-18, 2005, Proceedings}}},\ \bibinfo {series}
  {Lecture Notes in Computer Science}, Vol.\ \bibinfo {volume} {3621}\
  (\bibinfo {year} {2005})\BibitemShut {NoStop}%
\bibitem [{\citenamefont {Stevens}\ \emph {et~al.}(2017)\citenamefont
  {Stevens}, \citenamefont {Bursztein}, \citenamefont {Karpman}, \citenamefont
  {Albertini},\ and\ \citenamefont {Markov}}]{Stevens:2017:The}%
  \BibitemOpen
  \bibfield  {author} {\bibinfo {author} {\bibfnamefont {M.}~\bibnamefont
  {Stevens}}, \bibinfo {author} {\bibfnamefont {E.}~\bibnamefont {Bursztein}},
  \bibinfo {author} {\bibfnamefont {P.}~\bibnamefont {Karpman}}, \bibinfo
  {author} {\bibfnamefont {A.}~\bibnamefont {Albertini}},\ and\ \bibinfo
  {author} {\bibfnamefont {Y.}~\bibnamefont {Markov}},\ }\href@noop {}
  {\bibinfo {title} {The first collision for full sha-1}} (\bibinfo {year}
  {2017})\BibitemShut {NoStop}%
\bibitem [{\citenamefont {Kleinjung}\ \emph {et~al.}(2010)\citenamefont
  {Kleinjung}, \citenamefont {Aoki}, \citenamefont {Franke}, \citenamefont
  {Lenstra}, \citenamefont {Thom{\'e}}, \citenamefont {Bos}, \citenamefont
  {Gaudry}, \citenamefont {Kruppa}, \citenamefont {Montgomery}, \citenamefont
  {Osvik}, \citenamefont {te~Riele}, \citenamefont {Timofeev},\ and\
  \citenamefont {Zimmermann}}]{kleinjung2010factorization}%
  \BibitemOpen
  \bibfield  {author} {\bibinfo {author} {\bibfnamefont {T.}~\bibnamefont
  {Kleinjung}}, \bibinfo {author} {\bibfnamefont {K.}~\bibnamefont {Aoki}},
  \bibinfo {author} {\bibfnamefont {J.}~\bibnamefont {Franke}}, \bibinfo
  {author} {\bibfnamefont {A.~K.}\ \bibnamefont {Lenstra}}, \bibinfo {author}
  {\bibfnamefont {E.}~\bibnamefont {Thom{\'e}}}, \bibinfo {author}
  {\bibfnamefont {J.~W.}\ \bibnamefont {Bos}}, \bibinfo {author} {\bibfnamefont
  {P.}~\bibnamefont {Gaudry}}, \bibinfo {author} {\bibfnamefont
  {A.}~\bibnamefont {Kruppa}}, \bibinfo {author} {\bibfnamefont {P.~L.}\
  \bibnamefont {Montgomery}}, \bibinfo {author} {\bibfnamefont {D.~A.}\
  \bibnamefont {Osvik}}, \bibinfo {author} {\bibfnamefont {H.}~\bibnamefont
  {te~Riele}}, \bibinfo {author} {\bibfnamefont {A.}~\bibnamefont {Timofeev}},\
  and\ \bibinfo {author} {\bibfnamefont {P.}~\bibnamefont {Zimmermann}},\
  }\bibfield  {title} {\bibinfo {title} {Factorization of a 768-bit rsa
  modulus},\ }in\ \href@noop {} {\emph {\bibinfo {booktitle} {Advances in
  Cryptology -- CRYPTO 2010}}}\ (\bibinfo {year} {2010})\BibitemShut {NoStop}%
\bibitem [{\citenamefont {Kleinjung}\ \emph {et~al.}(2017)\citenamefont
  {Kleinjung}, \citenamefont {Diem}, \citenamefont {Lenstra}, \citenamefont
  {Priplata},\ and\ \citenamefont {Stahlke}}]{kleinjung2017computation}%
  \BibitemOpen
  \bibfield  {author} {\bibinfo {author} {\bibfnamefont {T.}~\bibnamefont
  {Kleinjung}}, \bibinfo {author} {\bibfnamefont {C.}~\bibnamefont {Diem}},
  \bibinfo {author} {\bibfnamefont {A.~K.}\ \bibnamefont {Lenstra}}, \bibinfo
  {author} {\bibfnamefont {C.}~\bibnamefont {Priplata}},\ and\ \bibinfo
  {author} {\bibfnamefont {C.}~\bibnamefont {Stahlke}},\ }\href@noop {}
  {\bibinfo {title} {Computation of a 768-bit prime field discrete logarithm}}
  (\bibinfo {year} {2017})\BibitemShut {NoStop}%
\bibitem [{\citenamefont {Boudot}\ \emph {et~al.}(2020)\citenamefont {Boudot},
  \citenamefont {Gaudry}, \citenamefont {Guillevic}, \citenamefont {Heninger},
  \citenamefont {Thom{\'e}},\ and\ \citenamefont
  {Zimmermann}}]{boudot2020comparing}%
  \BibitemOpen
  \bibfield  {author} {\bibinfo {author} {\bibfnamefont {F.}~\bibnamefont
  {Boudot}}, \bibinfo {author} {\bibfnamefont {P.}~\bibnamefont {Gaudry}},
  \bibinfo {author} {\bibfnamefont {A.}~\bibnamefont {Guillevic}}, \bibinfo
  {author} {\bibfnamefont {N.}~\bibnamefont {Heninger}}, \bibinfo {author}
  {\bibfnamefont {E.}~\bibnamefont {Thom{\'e}}},\ and\ \bibinfo {author}
  {\bibfnamefont {P.}~\bibnamefont {Zimmermann}},\ }\href@noop {} {\bibinfo
  {title} {Comparing the difficulty of factorization and discrete logarithm: a
  240-digit experiment}} (\bibinfo {year} {2020})\BibitemShut {NoStop}%
\bibitem [{\citenamefont {Shor}(1994)}]{shor1994algorithms}%
  \BibitemOpen
  \bibfield  {author} {\bibinfo {author} {\bibfnamefont {P.~W.}\ \bibnamefont
  {Shor}},\ }\href@noop {} {\bibinfo {title} {Algorithms for quantum
  computation: discrete logarithms and factoring}} (\bibinfo {year}
  {1994})\BibitemShut {NoStop}%
\bibitem [{\citenamefont {Shannon}(1949)}]{shannon1949communication}%
  \BibitemOpen
  \bibfield  {author} {\bibinfo {author} {\bibfnamefont {C.~E.}\ \bibnamefont
  {Shannon}},\ }\bibfield  {title} {\bibinfo {title} {Communication theory of
  secrecy systems},\ }\href@noop {} {\bibfield  {journal} {\bibinfo  {journal}
  {The Bell system technical journal}\ }\textbf {\bibinfo {volume} {28}},\
  \bibinfo {pages} {656} (\bibinfo {year} {1949})}\BibitemShut {NoStop}%
\bibitem [{\citenamefont {Bennett}\ and\ \citenamefont
  {Brassard}(2014)}]{bennett2014quantum}%
  \BibitemOpen
  \bibfield  {author} {\bibinfo {author} {\bibfnamefont {C.~H.}\ \bibnamefont
  {Bennett}}\ and\ \bibinfo {author} {\bibfnamefont {G.}~\bibnamefont
  {Brassard}},\ }\bibfield  {title} {\bibinfo {title} {Quantum cryptography:
  Public key distribution and coin tossing},\ }\href@noop {} {\bibfield
  {journal} {\bibinfo  {journal} {Theor. Comput. Sci.}\ }\textbf {\bibinfo
  {volume} {560}},\ \bibinfo {pages} {7} (\bibinfo {year} {2014})}\BibitemShut
  {NoStop}%
\bibitem [{\citenamefont {Chen}\ \emph {et~al.}(2021)\citenamefont {Chen},
  \citenamefont {Zhang}, \citenamefont {Chen}, \citenamefont {Cai},
  \citenamefont {Liao}, \citenamefont {Zhang}, \citenamefont {Chen},
  \citenamefont {Yin}, \citenamefont {Ren}, \citenamefont {Chen} \emph
  {et~al.}}]{chen2021integrated}%
  \BibitemOpen
  \bibfield  {author} {\bibinfo {author} {\bibfnamefont {Y.-A.}\ \bibnamefont
  {Chen}}, \bibinfo {author} {\bibfnamefont {Q.}~\bibnamefont {Zhang}},
  \bibinfo {author} {\bibfnamefont {T.-Y.}\ \bibnamefont {Chen}}, \bibinfo
  {author} {\bibfnamefont {W.-Q.}\ \bibnamefont {Cai}}, \bibinfo {author}
  {\bibfnamefont {S.-K.}\ \bibnamefont {Liao}}, \bibinfo {author}
  {\bibfnamefont {J.}~\bibnamefont {Zhang}}, \bibinfo {author} {\bibfnamefont
  {K.}~\bibnamefont {Chen}}, \bibinfo {author} {\bibfnamefont {J.}~\bibnamefont
  {Yin}}, \bibinfo {author} {\bibfnamefont {J.-G.}\ \bibnamefont {Ren}},
  \bibinfo {author} {\bibfnamefont {Z.}~\bibnamefont {Chen}}, \emph {et~al.},\
  }\bibfield  {title} {\bibinfo {title} {An integrated space-to-ground quantum
  communication network over 4,600 kilometres},\ }\href@noop {} {\bibfield
  {journal} {\bibinfo  {journal} {Nature}\ }\textbf {\bibinfo {volume} {589}},\
  \bibinfo {pages} {214} (\bibinfo {year} {2021})}\BibitemShut {NoStop}%
\bibitem [{\citenamefont {Liu}\ \emph {et~al.}(2020)\citenamefont {Liu},
  \citenamefont {Tian}, \citenamefont {Gu}, \citenamefont {Fan}, \citenamefont
  {Ni}, \citenamefont {Yang}, \citenamefont {Zhang}, \citenamefont {Hu},
  \citenamefont {Guo}, \citenamefont {Cao} \emph {et~al.}}]{liu2020drone}%
  \BibitemOpen
  \bibfield  {author} {\bibinfo {author} {\bibfnamefont {H.-Y.}\ \bibnamefont
  {Liu}}, \bibinfo {author} {\bibfnamefont {X.-H.}\ \bibnamefont {Tian}},
  \bibinfo {author} {\bibfnamefont {C.}~\bibnamefont {Gu}}, \bibinfo {author}
  {\bibfnamefont {P.}~\bibnamefont {Fan}}, \bibinfo {author} {\bibfnamefont
  {X.}~\bibnamefont {Ni}}, \bibinfo {author} {\bibfnamefont {R.}~\bibnamefont
  {Yang}}, \bibinfo {author} {\bibfnamefont {J.-N.}\ \bibnamefont {Zhang}},
  \bibinfo {author} {\bibfnamefont {M.}~\bibnamefont {Hu}}, \bibinfo {author}
  {\bibfnamefont {J.}~\bibnamefont {Guo}}, \bibinfo {author} {\bibfnamefont
  {X.}~\bibnamefont {Cao}}, \emph {et~al.},\ }\bibfield  {title} {\bibinfo
  {title} {Drone-based entanglement distribution towards mobile quantum
  networks},\ }\href@noop {} {\bibfield  {journal} {\bibinfo  {journal}
  {National Science Review}\ }\textbf {\bibinfo {volume} {7}},\ \bibinfo
  {pages} {921} (\bibinfo {year} {2020})}\BibitemShut {NoStop}%
\bibitem [{\citenamefont {Kwek}\ \emph {et~al.}(2021)\citenamefont {Kwek},
  \citenamefont {Cao}, \citenamefont {Luo}, \citenamefont {Wang}, \citenamefont
  {Sun}, \citenamefont {Wang},\ and\ \citenamefont {Liu}}]{kwek2021chip}%
  \BibitemOpen
  \bibfield  {author} {\bibinfo {author} {\bibfnamefont {L.-C.}\ \bibnamefont
  {Kwek}}, \bibinfo {author} {\bibfnamefont {L.}~\bibnamefont {Cao}}, \bibinfo
  {author} {\bibfnamefont {W.}~\bibnamefont {Luo}}, \bibinfo {author}
  {\bibfnamefont {Y.}~\bibnamefont {Wang}}, \bibinfo {author} {\bibfnamefont
  {S.}~\bibnamefont {Sun}}, \bibinfo {author} {\bibfnamefont {X.}~\bibnamefont
  {Wang}},\ and\ \bibinfo {author} {\bibfnamefont {A.~Q.}\ \bibnamefont
  {Liu}},\ }\bibfield  {title} {\bibinfo {title} {Chip-based quantum key
  distribution},\ }\href@noop {} {\bibfield  {journal} {\bibinfo  {journal}
  {AAPPS Bulletin}\ }\textbf {\bibinfo {volume} {31}},\ \bibinfo {pages} {1}
  (\bibinfo {year} {2021})}\BibitemShut {NoStop}%
\bibitem [{\citenamefont {Zhang}\ and\ \citenamefont
  {Ni}(2020)}]{zhang2020design}%
  \BibitemOpen
  \bibfield  {author} {\bibinfo {author} {\bibfnamefont {Y.}~\bibnamefont
  {Zhang}}\ and\ \bibinfo {author} {\bibfnamefont {Q.}~\bibnamefont {Ni}},\
  }\bibfield  {title} {\bibinfo {title} {Design and analysis of random multiple
  access quantum key distribution},\ }\href@noop {} {\bibfield  {journal}
  {\bibinfo  {journal} {Quantum Engineering}\ }\textbf {\bibinfo {volume}
  {2}},\ \bibinfo {pages} {e31} (\bibinfo {year} {2020})}\BibitemShut {NoStop}%
\bibitem [{\citenamefont {Guo}\ \emph {et~al.}(2021)\citenamefont {Guo},
  \citenamefont {Li}, \citenamefont {Yu},\ and\ \citenamefont
  {Zhang}}]{guo2021toward}%
  \BibitemOpen
  \bibfield  {author} {\bibinfo {author} {\bibfnamefont {H.}~\bibnamefont
  {Guo}}, \bibinfo {author} {\bibfnamefont {Z.}~\bibnamefont {Li}}, \bibinfo
  {author} {\bibfnamefont {S.}~\bibnamefont {Yu}},\ and\ \bibinfo {author}
  {\bibfnamefont {Y.}~\bibnamefont {Zhang}},\ }\bibfield  {title} {\bibinfo
  {title} {Toward practical quantum key distribution using telecom
  components},\ }\href@noop {} {\bibfield  {journal} {\bibinfo  {journal}
  {Fundamental Research}\ }\textbf {\bibinfo {volume} {1}},\ \bibinfo {pages}
  {96} (\bibinfo {year} {2021})}\BibitemShut {NoStop}%
\bibitem [{\citenamefont {Long}\ and\ \citenamefont
  {Liu}(2002)}]{long2002theoretically}%
  \BibitemOpen
  \bibfield  {author} {\bibinfo {author} {\bibfnamefont {G.-L.}\ \bibnamefont
  {Long}}\ and\ \bibinfo {author} {\bibfnamefont {X.-S.}\ \bibnamefont {Liu}},\
  }\bibfield  {title} {\bibinfo {title} {Theoretically efficient high-capacity
  quantum-key-distribution scheme},\ }\href@noop {} {\bibfield  {journal}
  {\bibinfo  {journal} {Phys. Rev. A}\ }\textbf {\bibinfo {volume} {65}},\
  \bibinfo {pages} {032302} (\bibinfo {year} {2002})}\BibitemShut {NoStop}%
\bibitem [{\citenamefont {Qi}\ \emph {et~al.}(2021)\citenamefont {Qi},
  \citenamefont {Li}, \citenamefont {Huang}, \citenamefont {Feng},
  \citenamefont {Zheng},\ and\ \citenamefont {Chen}}]{qi202115}%
  \BibitemOpen
  \bibfield  {author} {\bibinfo {author} {\bibfnamefont {Z.}~\bibnamefont
  {Qi}}, \bibinfo {author} {\bibfnamefont {Y.}~\bibnamefont {Li}}, \bibinfo
  {author} {\bibfnamefont {Y.}~\bibnamefont {Huang}}, \bibinfo {author}
  {\bibfnamefont {J.}~\bibnamefont {Feng}}, \bibinfo {author} {\bibfnamefont
  {Y.}~\bibnamefont {Zheng}},\ and\ \bibinfo {author} {\bibfnamefont
  {X.}~\bibnamefont {Chen}},\ }\bibfield  {title} {\bibinfo {title} {A 15-user
  quantum secure direct communication network},\ }\href@noop {} {\bibfield
  {journal} {\bibinfo  {journal} {Light Sci. Appl.}\ }\textbf {\bibinfo
  {volume} {10}},\ \bibinfo {pages} {183} (\bibinfo {year} {2021})}\BibitemShut
  {NoStop}%
\bibitem [{\citenamefont {Sheng}\ \emph {et~al.}(2022)\citenamefont {Sheng},
  \citenamefont {Zhou},\ and\ \citenamefont {Long}}]{Sheng:2022:One}%
  \BibitemOpen
  \bibfield  {author} {\bibinfo {author} {\bibfnamefont {Y.-B.}\ \bibnamefont
  {Sheng}}, \bibinfo {author} {\bibfnamefont {L.}~\bibnamefont {Zhou}},\ and\
  \bibinfo {author} {\bibfnamefont {G.-L.}\ \bibnamefont {Long}},\ }\bibfield
  {title} {\bibinfo {title} {One-step quantum secure direct communication},\
  }\href@noop {} {\bibfield  {journal} {\bibinfo  {journal} {Sci. Bull.}\
  }\textbf {\bibinfo {volume} {67}},\ \bibinfo {pages} {367} (\bibinfo {year}
  {2022})}\BibitemShut {NoStop}%
\bibitem [{\citenamefont {Lyubashevsky}(2021)}]{lyubashevsky2021lattice}%
  \BibitemOpen
  \bibfield  {author} {\bibinfo {author} {\bibfnamefont {V.}~\bibnamefont
  {Lyubashevsky}},\ }\bibfield  {title} {\bibinfo {title} {Lattice-based
  digital signatures},\ }\href@noop {} {\bibfield  {journal} {\bibinfo
  {journal} {National Science Review}\ }\textbf {\bibinfo {volume} {8}},\
  \bibinfo {pages} {nwab077} (\bibinfo {year} {2021})}\BibitemShut {NoStop}%
\bibitem [{\citenamefont {Gottesman}\ and\ \citenamefont
  {Chuang}(2001)}]{gottesman2001quantum}%
  \BibitemOpen
  \bibfield  {author} {\bibinfo {author} {\bibfnamefont {D.}~\bibnamefont
  {Gottesman}}\ and\ \bibinfo {author} {\bibfnamefont {I.}~\bibnamefont
  {Chuang}},\ }\bibfield  {title} {\bibinfo {title} {Quantum digital
  signatures},\ }\href@noop {} {\bibfield  {journal} {\bibinfo  {journal}
  {arXiv preprint quant-ph/0105032}\ } (\bibinfo {year} {2001})}\BibitemShut
  {NoStop}%
\bibitem [{\citenamefont {Clarke}\ \emph {et~al.}(2012)\citenamefont {Clarke},
  \citenamefont {Collins}, \citenamefont {Dunjko}, \citenamefont {Andersson},
  \citenamefont {Jeffers},\ and\ \citenamefont
  {Buller}}]{clarke2012experimental}%
  \BibitemOpen
  \bibfield  {author} {\bibinfo {author} {\bibfnamefont {P.~J.}\ \bibnamefont
  {Clarke}}, \bibinfo {author} {\bibfnamefont {R.~J.}\ \bibnamefont {Collins}},
  \bibinfo {author} {\bibfnamefont {V.}~\bibnamefont {Dunjko}}, \bibinfo
  {author} {\bibfnamefont {E.}~\bibnamefont {Andersson}}, \bibinfo {author}
  {\bibfnamefont {J.}~\bibnamefont {Jeffers}},\ and\ \bibinfo {author}
  {\bibfnamefont {G.~S.}\ \bibnamefont {Buller}},\ }\bibfield  {title}
  {\bibinfo {title} {Experimental demonstration of quantum digital signatures
  using phase-encoded coherent states of light},\ }\href@noop {} {\bibfield
  {journal} {\bibinfo  {journal} {Nature Commun.}\ }\textbf {\bibinfo {volume}
  {3}},\ \bibinfo {pages} {1174} (\bibinfo {year} {2012})}\BibitemShut
  {NoStop}%
\bibitem [{\citenamefont {Dunjko}\ \emph {et~al.}(2014)\citenamefont {Dunjko},
  \citenamefont {Wallden},\ and\ \citenamefont
  {Andersson}}]{dunjko2014quantum}%
  \BibitemOpen
  \bibfield  {author} {\bibinfo {author} {\bibfnamefont {V.}~\bibnamefont
  {Dunjko}}, \bibinfo {author} {\bibfnamefont {P.}~\bibnamefont {Wallden}},\
  and\ \bibinfo {author} {\bibfnamefont {E.}~\bibnamefont {Andersson}},\
  }\bibfield  {title} {\bibinfo {title} {Quantum digital signatures without
  quantum memory},\ }\href@noop {} {\bibfield  {journal} {\bibinfo  {journal}
  {Phys. Rev. Lett.}\ }\textbf {\bibinfo {volume} {112}},\ \bibinfo {pages}
  {040502} (\bibinfo {year} {2014})}\BibitemShut {NoStop}%
\bibitem [{\citenamefont {Yin}\ \emph {et~al.}(2016)\citenamefont {Yin},
  \citenamefont {Fu},\ and\ \citenamefont {Chen}}]{Yin:practical:2016}%
  \BibitemOpen
  \bibfield  {author} {\bibinfo {author} {\bibfnamefont {H.-L.}\ \bibnamefont
  {Yin}}, \bibinfo {author} {\bibfnamefont {Y.}~\bibnamefont {Fu}},\ and\
  \bibinfo {author} {\bibfnamefont {Z.-B.}\ \bibnamefont {Chen}},\ }\bibfield
  {title} {\bibinfo {title} {Practical quantum digital signature},\ }\href@noop
  {} {\bibfield  {journal} {\bibinfo  {journal} {Phys. Rev. A}\ }\textbf
  {\bibinfo {volume} {93}},\ \bibinfo {pages} {032316} (\bibinfo {year}
  {2016})}\BibitemShut {NoStop}%
\bibitem [{\citenamefont {Amiri}\ \emph {et~al.}(2016)\citenamefont {Amiri},
  \citenamefont {Wallden}, \citenamefont {Kent},\ and\ \citenamefont
  {Andersson}}]{Amiri:Secure:2016}%
  \BibitemOpen
  \bibfield  {author} {\bibinfo {author} {\bibfnamefont {R.}~\bibnamefont
  {Amiri}}, \bibinfo {author} {\bibfnamefont {P.}~\bibnamefont {Wallden}},
  \bibinfo {author} {\bibfnamefont {A.}~\bibnamefont {Kent}},\ and\ \bibinfo
  {author} {\bibfnamefont {E.}~\bibnamefont {Andersson}},\ }\bibfield  {title}
  {\bibinfo {title} {Secure quantum signatures using insecure quantum
  channels},\ }\href@noop {} {\bibfield  {journal} {\bibinfo  {journal} {Phys.
  Rev. A}\ }\textbf {\bibinfo {volume} {93}},\ \bibinfo {pages} {032325}
  (\bibinfo {year} {2016})}\BibitemShut {NoStop}%
\bibitem [{\citenamefont {Lu}\ \emph {et~al.}(2021)\citenamefont {Lu},
  \citenamefont {Cao}, \citenamefont {Weng}, \citenamefont {Gu}, \citenamefont
  {Xie}, \citenamefont {Zhou}, \citenamefont {Yin},\ and\ \citenamefont
  {Chen}}]{lu2021efficient}%
  \BibitemOpen
  \bibfield  {author} {\bibinfo {author} {\bibfnamefont {Y.-S.}\ \bibnamefont
  {Lu}}, \bibinfo {author} {\bibfnamefont {X.-Y.}\ \bibnamefont {Cao}},
  \bibinfo {author} {\bibfnamefont {C.-X.}\ \bibnamefont {Weng}}, \bibinfo
  {author} {\bibfnamefont {J.}~\bibnamefont {Gu}}, \bibinfo {author}
  {\bibfnamefont {Y.-M.}\ \bibnamefont {Xie}}, \bibinfo {author} {\bibfnamefont
  {M.-G.}\ \bibnamefont {Zhou}}, \bibinfo {author} {\bibfnamefont {H.-L.}\
  \bibnamefont {Yin}},\ and\ \bibinfo {author} {\bibfnamefont {Z.-B.}\
  \bibnamefont {Chen}},\ }\bibfield  {title} {\bibinfo {title} {Efficient
  quantum digital signatures without symmetrization step},\ }\href@noop {}
  {\bibfield  {journal} {\bibinfo  {journal} {Opt. Express}\ }\textbf {\bibinfo
  {volume} {29}},\ \bibinfo {pages} {10162} (\bibinfo {year}
  {2021})}\BibitemShut {NoStop}%
\bibitem [{\citenamefont {Weng}\ \emph {et~al.}(2021)\citenamefont {Weng},
  \citenamefont {Lu}, \citenamefont {Gao}, \citenamefont {Xie}, \citenamefont
  {Gu}, \citenamefont {Li}, \citenamefont {Li}, \citenamefont {Yin},\ and\
  \citenamefont {Chen}}]{weng2021secure}%
  \BibitemOpen
  \bibfield  {author} {\bibinfo {author} {\bibfnamefont {C.-X.}\ \bibnamefont
  {Weng}}, \bibinfo {author} {\bibfnamefont {Y.-S.}\ \bibnamefont {Lu}},
  \bibinfo {author} {\bibfnamefont {R.-Q.}\ \bibnamefont {Gao}}, \bibinfo
  {author} {\bibfnamefont {Y.-M.}\ \bibnamefont {Xie}}, \bibinfo {author}
  {\bibfnamefont {J.}~\bibnamefont {Gu}}, \bibinfo {author} {\bibfnamefont
  {C.-L.}\ \bibnamefont {Li}}, \bibinfo {author} {\bibfnamefont {B.-H.}\
  \bibnamefont {Li}}, \bibinfo {author} {\bibfnamefont {H.-L.}\ \bibnamefont
  {Yin}},\ and\ \bibinfo {author} {\bibfnamefont {Z.-B.}\ \bibnamefont
  {Chen}},\ }\bibfield  {title} {\bibinfo {title} {Secure and practical
  multiparty quantum digital signatures},\ }\href@noop {} {\bibfield  {journal}
  {\bibinfo  {journal} {Opt. Express}\ }\textbf {\bibinfo {volume} {29}},\
  \bibinfo {pages} {27661} (\bibinfo {year} {2021})}\BibitemShut {NoStop}%
\bibitem [{\citenamefont {Yin}\ \emph {et~al.}(2017{\natexlab{a}})\citenamefont
  {Yin}, \citenamefont {Fu}, \citenamefont {Liu}, \citenamefont {Tang},
  \citenamefont {Wang}, \citenamefont {You}, \citenamefont {Zhang},
  \citenamefont {Chen}, \citenamefont {Wang}, \citenamefont {Zhang},
  \citenamefont {Chen}, \citenamefont {Chen},\ and\ \citenamefont
  {Pan}}]{Yin:2017:Experimental}%
  \BibitemOpen
  \bibfield  {author} {\bibinfo {author} {\bibfnamefont {H.-L.}\ \bibnamefont
  {Yin}}, \bibinfo {author} {\bibfnamefont {Y.}~\bibnamefont {Fu}}, \bibinfo
  {author} {\bibfnamefont {H.}~\bibnamefont {Liu}}, \bibinfo {author}
  {\bibfnamefont {Q.-J.}\ \bibnamefont {Tang}}, \bibinfo {author}
  {\bibfnamefont {J.}~\bibnamefont {Wang}}, \bibinfo {author} {\bibfnamefont
  {L.-X.}\ \bibnamefont {You}}, \bibinfo {author} {\bibfnamefont {W.-J.}\
  \bibnamefont {Zhang}}, \bibinfo {author} {\bibfnamefont {S.-J.}\ \bibnamefont
  {Chen}}, \bibinfo {author} {\bibfnamefont {Z.}~\bibnamefont {Wang}}, \bibinfo
  {author} {\bibfnamefont {Q.}~\bibnamefont {Zhang}}, \bibinfo {author}
  {\bibfnamefont {T.-Y.}\ \bibnamefont {Chen}}, \bibinfo {author}
  {\bibfnamefont {Z.-B.}\ \bibnamefont {Chen}},\ and\ \bibinfo {author}
  {\bibfnamefont {J.-W.}\ \bibnamefont {Pan}},\ }\bibfield  {title} {\bibinfo
  {title} {Experimental quantum digital signature over 102 km},\ }\href@noop {}
  {\bibfield  {journal} {\bibinfo  {journal} {Phys. Rev. A}\ }\textbf {\bibinfo
  {volume} {95}},\ \bibinfo {pages} {032334} (\bibinfo {year}
  {2017}{\natexlab{a}})}\BibitemShut {NoStop}%
\bibitem [{\citenamefont {Yin}\ \emph {et~al.}(2017{\natexlab{b}})\citenamefont
  {Yin}, \citenamefont {Wang}, \citenamefont {Tang}, \citenamefont {Zhao},
  \citenamefont {Liu}, \citenamefont {Sun}, \citenamefont {Zhang},
  \citenamefont {Li}, \citenamefont {Puthoor}, \citenamefont {You},
  \citenamefont {Andersson}, \citenamefont {Wang}, \citenamefont {Liu},
  \citenamefont {Jiang}, \citenamefont {Ma}, \citenamefont {Zhang},
  \citenamefont {Curty}, \citenamefont {Chen},\ and\ \citenamefont
  {Pan}}]{Yin:2017:Exp}%
  \BibitemOpen
  \bibfield  {author} {\bibinfo {author} {\bibfnamefont {H.-L.}\ \bibnamefont
  {Yin}}, \bibinfo {author} {\bibfnamefont {W.-L.}\ \bibnamefont {Wang}},
  \bibinfo {author} {\bibfnamefont {Y.-L.}\ \bibnamefont {Tang}}, \bibinfo
  {author} {\bibfnamefont {Q.}~\bibnamefont {Zhao}}, \bibinfo {author}
  {\bibfnamefont {H.}~\bibnamefont {Liu}}, \bibinfo {author} {\bibfnamefont
  {X.-X.}\ \bibnamefont {Sun}}, \bibinfo {author} {\bibfnamefont {W.-J.}\
  \bibnamefont {Zhang}}, \bibinfo {author} {\bibfnamefont {H.}~\bibnamefont
  {Li}}, \bibinfo {author} {\bibfnamefont {I.~V.}\ \bibnamefont {Puthoor}},
  \bibinfo {author} {\bibfnamefont {L.-X.}\ \bibnamefont {You}}, \bibinfo
  {author} {\bibfnamefont {E.}~\bibnamefont {Andersson}}, \bibinfo {author}
  {\bibfnamefont {Z.}~\bibnamefont {Wang}}, \bibinfo {author} {\bibfnamefont
  {Y.}~\bibnamefont {Liu}}, \bibinfo {author} {\bibfnamefont {X.}~\bibnamefont
  {Jiang}}, \bibinfo {author} {\bibfnamefont {X.}~\bibnamefont {Ma}}, \bibinfo
  {author} {\bibfnamefont {Q.}~\bibnamefont {Zhang}}, \bibinfo {author}
  {\bibfnamefont {M.}~\bibnamefont {Curty}}, \bibinfo {author} {\bibfnamefont
  {T.-Y.}\ \bibnamefont {Chen}},\ and\ \bibinfo {author} {\bibfnamefont
  {J.-W.}\ \bibnamefont {Pan}},\ }\bibfield  {title} {\bibinfo {title}
  {Experimental measurement-device-independent quantum digital signatures over
  a metropolitan network},\ }\href@noop {} {\bibfield  {journal} {\bibinfo
  {journal} {Phys. Rev. A}\ }\textbf {\bibinfo {volume} {95}},\ \bibinfo
  {pages} {042338} (\bibinfo {year} {2017}{\natexlab{b}})}\BibitemShut
  {NoStop}%
\bibitem [{\citenamefont {Roberts}\ \emph {et~al.}(2017)\citenamefont
  {Roberts}, \citenamefont {Lucamarini}, \citenamefont {Yuan}, \citenamefont
  {Dynes}, \citenamefont {Comandar}, \citenamefont {Sharpe}, \citenamefont
  {Shields}, \citenamefont {Curty}, \citenamefont {Puthoor},\ and\
  \citenamefont {Andersson}}]{roberts2017experimental}%
  \BibitemOpen
  \bibfield  {author} {\bibinfo {author} {\bibfnamefont {G.}~\bibnamefont
  {Roberts}}, \bibinfo {author} {\bibfnamefont {M.}~\bibnamefont {Lucamarini}},
  \bibinfo {author} {\bibfnamefont {Z.}~\bibnamefont {Yuan}}, \bibinfo {author}
  {\bibfnamefont {J.}~\bibnamefont {Dynes}}, \bibinfo {author} {\bibfnamefont
  {L.}~\bibnamefont {Comandar}}, \bibinfo {author} {\bibfnamefont
  {A.}~\bibnamefont {Sharpe}}, \bibinfo {author} {\bibfnamefont
  {A.}~\bibnamefont {Shields}}, \bibinfo {author} {\bibfnamefont
  {M.}~\bibnamefont {Curty}}, \bibinfo {author} {\bibfnamefont
  {I.}~\bibnamefont {Puthoor}},\ and\ \bibinfo {author} {\bibfnamefont
  {E.}~\bibnamefont {Andersson}},\ }\bibfield  {title} {\bibinfo {title}
  {Experimental measurement-device-independent quantum digital signatures},\
  }\href@noop {} {\bibfield  {journal} {\bibinfo  {journal} {Nature Commun.}\
  }\textbf {\bibinfo {volume} {8}},\ \bibinfo {pages} {1098} (\bibinfo {year}
  {2017})}\BibitemShut {NoStop}%
\bibitem [{\citenamefont {An}\ \emph {et~al.}(2019)\citenamefont {An},
  \citenamefont {Zhang}, \citenamefont {Zhang}, \citenamefont {Chen},
  \citenamefont {Wang}, \citenamefont {Yin}, \citenamefont {Wang},
  \citenamefont {He}, \citenamefont {Hao}, \citenamefont {Liu} \emph
  {et~al.}}]{an2019practical}%
  \BibitemOpen
  \bibfield  {author} {\bibinfo {author} {\bibfnamefont {X.-B.}\ \bibnamefont
  {An}}, \bibinfo {author} {\bibfnamefont {H.}~\bibnamefont {Zhang}}, \bibinfo
  {author} {\bibfnamefont {C.-M.}\ \bibnamefont {Zhang}}, \bibinfo {author}
  {\bibfnamefont {W.}~\bibnamefont {Chen}}, \bibinfo {author} {\bibfnamefont
  {S.}~\bibnamefont {Wang}}, \bibinfo {author} {\bibfnamefont {Z.-Q.}\
  \bibnamefont {Yin}}, \bibinfo {author} {\bibfnamefont {Q.}~\bibnamefont
  {Wang}}, \bibinfo {author} {\bibfnamefont {D.-Y.}\ \bibnamefont {He}},
  \bibinfo {author} {\bibfnamefont {P.-L.}\ \bibnamefont {Hao}}, \bibinfo
  {author} {\bibfnamefont {S.-F.}\ \bibnamefont {Liu}}, \emph {et~al.},\
  }\bibfield  {title} {\bibinfo {title} {Practical quantum digital signature
  with a gigahertz bb84 quantum key distribution system},\ }\href@noop {}
  {\bibfield  {journal} {\bibinfo  {journal} {Opt. Lett.}\ }\textbf {\bibinfo
  {volume} {44}},\ \bibinfo {pages} {139} (\bibinfo {year} {2019})}\BibitemShut
  {NoStop}%
\bibitem [{\citenamefont {Richter}\ \emph {et~al.}(2021)\citenamefont
  {Richter}, \citenamefont {Thornton}, \citenamefont {Khan}, \citenamefont
  {Scott}, \citenamefont {Jaksch}, \citenamefont {Vogl}, \citenamefont
  {Stiller}, \citenamefont {Leuchs}, \citenamefont {Marquardt},\ and\
  \citenamefont {Korolkova}}]{Richter2021Agile}%
  \BibitemOpen
  \bibfield  {author} {\bibinfo {author} {\bibfnamefont {S.}~\bibnamefont
  {Richter}}, \bibinfo {author} {\bibfnamefont {M.}~\bibnamefont {Thornton}},
  \bibinfo {author} {\bibfnamefont {I.}~\bibnamefont {Khan}}, \bibinfo {author}
  {\bibfnamefont {H.}~\bibnamefont {Scott}}, \bibinfo {author} {\bibfnamefont
  {K.}~\bibnamefont {Jaksch}}, \bibinfo {author} {\bibfnamefont
  {U.}~\bibnamefont {Vogl}}, \bibinfo {author} {\bibfnamefont {B.}~\bibnamefont
  {Stiller}}, \bibinfo {author} {\bibfnamefont {G.}~\bibnamefont {Leuchs}},
  \bibinfo {author} {\bibfnamefont {C.}~\bibnamefont {Marquardt}},\ and\
  \bibinfo {author} {\bibfnamefont {N.}~\bibnamefont {Korolkova}},\ }\bibfield
  {title} {\bibinfo {title} {Agile and versatile quantum communication:
  Signatures and secrets},\ }\href@noop {} {\bibfield  {journal} {\bibinfo
  {journal} {Phys. Rev. X}\ }\textbf {\bibinfo {volume} {11}},\ \bibinfo
  {pages} {011038} (\bibinfo {year} {2021})}\BibitemShut {NoStop}%
\bibitem [{\citenamefont {Wang}\ \emph {et~al.}(2015)\citenamefont {Wang},
  \citenamefont {Cai}, \citenamefont {Ren},\ and\ \citenamefont
  {Zhang}}]{wang2015security}%
  \BibitemOpen
  \bibfield  {author} {\bibinfo {author} {\bibfnamefont {T.-Y.}\ \bibnamefont
  {Wang}}, \bibinfo {author} {\bibfnamefont {X.-Q.}\ \bibnamefont {Cai}},
  \bibinfo {author} {\bibfnamefont {Y.-L.}\ \bibnamefont {Ren}},\ and\ \bibinfo
  {author} {\bibfnamefont {R.-L.}\ \bibnamefont {Zhang}},\ }\bibfield  {title}
  {\bibinfo {title} {Security of quantum digital signatures for classical
  messages},\ }\href@noop {} {\bibfield  {journal} {\bibinfo  {journal} {Sci.
  Rep.}\ }\textbf {\bibinfo {volume} {5}},\ \bibinfo {pages} {9231} (\bibinfo
  {year} {2015})}\BibitemShut {NoStop}%
\bibitem [{\citenamefont {Roehsner}\ \emph {et~al.}(2021)\citenamefont
  {Roehsner}, \citenamefont {Kettlewell}, \citenamefont {Fitzsimons},\ and\
  \citenamefont {Walther}}]{roehsner2021probabilistic}%
  \BibitemOpen
  \bibfield  {author} {\bibinfo {author} {\bibfnamefont {M.-C.}\ \bibnamefont
  {Roehsner}}, \bibinfo {author} {\bibfnamefont {J.~A.}\ \bibnamefont
  {Kettlewell}}, \bibinfo {author} {\bibfnamefont {J.}~\bibnamefont
  {Fitzsimons}},\ and\ \bibinfo {author} {\bibfnamefont {P.}~\bibnamefont
  {Walther}},\ }\bibfield  {title} {\bibinfo {title} {Probabilistic one-time
  programs using quantum entanglement},\ }\href@noop {} {\bibfield  {journal}
  {\bibinfo  {journal} {npj Quantum Inf.}\ }\textbf {\bibinfo {volume} {7}},\
  \bibinfo {pages} {98} (\bibinfo {year} {2021})}\BibitemShut {NoStop}%
\bibitem [{\citenamefont {Carter}\ and\ \citenamefont
  {Wegman}(1979)}]{carter1979universal}%
  \BibitemOpen
  \bibfield  {author} {\bibinfo {author} {\bibfnamefont {J.~L.}\ \bibnamefont
  {Carter}}\ and\ \bibinfo {author} {\bibfnamefont {M.~N.}\ \bibnamefont
  {Wegman}},\ }\bibfield  {title} {\bibinfo {title} {Universal classes of hash
  functions},\ }\href@noop {} {\bibfield  {journal} {\bibinfo  {journal}
  {Journal of Computer and System Sciences}\ }\textbf {\bibinfo {volume}
  {18}},\ \bibinfo {pages} {143} (\bibinfo {year} {1979})}\BibitemShut
  {NoStop}%
\bibitem [{\citenamefont {Wang}\ \emph {et~al.}(2018)\citenamefont {Wang},
  \citenamefont {Yu},\ and\ \citenamefont {Hu}}]{wang2018TFlargemisalignment}%
  \BibitemOpen
  \bibfield  {author} {\bibinfo {author} {\bibfnamefont {X.-B.}\ \bibnamefont
  {Wang}}, \bibinfo {author} {\bibfnamefont {Z.-W.}\ \bibnamefont {Yu}},\ and\
  \bibinfo {author} {\bibfnamefont {X.-L.}\ \bibnamefont {Hu}},\ }\bibfield
  {title} {\bibinfo {title} {Twin-field quantum key distribution with large
  misalignment error},\ }\href@noop {} {\bibfield  {journal} {\bibinfo
  {journal} {Phys. Rev. A}\ }\textbf {\bibinfo {volume} {98}},\ \bibinfo
  {pages} {062323} (\bibinfo {year} {2018})}\BibitemShut {NoStop}%
\bibitem [{\citenamefont {Ma}\ \emph {et~al.}(2018)\citenamefont {Ma},
  \citenamefont {Zeng},\ and\ \citenamefont {Zhou}}]{ma2018phase}%
  \BibitemOpen
  \bibfield  {author} {\bibinfo {author} {\bibfnamefont {X.}~\bibnamefont
  {Ma}}, \bibinfo {author} {\bibfnamefont {P.}~\bibnamefont {Zeng}},\ and\
  \bibinfo {author} {\bibfnamefont {H.}~\bibnamefont {Zhou}},\ }\bibfield
  {title} {\bibinfo {title} {Phase-matching quantum key distribution},\
  }\href@noop {} {\bibfield  {journal} {\bibinfo  {journal} {Phys. Rev. X}\
  }\textbf {\bibinfo {volume} {8}},\ \bibinfo {pages} {031043} (\bibinfo {year}
  {2018})}\BibitemShut {NoStop}%
\bibitem [{\citenamefont {Lin}\ \emph {et~al.}(2019)\citenamefont {Lin},
  \citenamefont {Upadhyaya},\ and\ \citenamefont
  {L{\"u}tkenhaus}}]{lin2019asymptotic}%
  \BibitemOpen
  \bibfield  {author} {\bibinfo {author} {\bibfnamefont {J.}~\bibnamefont
  {Lin}}, \bibinfo {author} {\bibfnamefont {T.}~\bibnamefont {Upadhyaya}},\
  and\ \bibinfo {author} {\bibfnamefont {N.}~\bibnamefont {L{\"u}tkenhaus}},\
  }\bibfield  {title} {\bibinfo {title} {Asymptotic security analysis of
  discrete-modulated continuous-variable quantum key distribution},\
  }\href@noop {} {\bibfield  {journal} {\bibinfo  {journal} {Phys. Rev. X}\
  }\textbf {\bibinfo {volume} {9}},\ \bibinfo {pages} {041064} (\bibinfo {year}
  {2019})}\BibitemShut {NoStop}%
\bibitem [{\citenamefont {Lodewyck}\ \emph {et~al.}(2007)\citenamefont
  {Lodewyck}, \citenamefont {Bloch}, \citenamefont {Garc\'{\i}a-Patr\'on},
  \citenamefont {Fossier}, \citenamefont {Karpov}, \citenamefont {Diamanti},
  \citenamefont {Debuisschert}, \citenamefont {Cerf}, \citenamefont
  {Tualle-Brouri}, \citenamefont {McLaughlin},\ and\ \citenamefont
  {Grangier}}]{Lodewyck2007Quantum}%
  \BibitemOpen
  \bibfield  {author} {\bibinfo {author} {\bibfnamefont {J.}~\bibnamefont
  {Lodewyck}}, \bibinfo {author} {\bibfnamefont {M.}~\bibnamefont {Bloch}},
  \bibinfo {author} {\bibfnamefont {R.}~\bibnamefont {Garc\'{\i}a-Patr\'on}},
  \bibinfo {author} {\bibfnamefont {S.}~\bibnamefont {Fossier}}, \bibinfo
  {author} {\bibfnamefont {E.}~\bibnamefont {Karpov}}, \bibinfo {author}
  {\bibfnamefont {E.}~\bibnamefont {Diamanti}}, \bibinfo {author}
  {\bibfnamefont {T.}~\bibnamefont {Debuisschert}}, \bibinfo {author}
  {\bibfnamefont {N.~J.}\ \bibnamefont {Cerf}}, \bibinfo {author}
  {\bibfnamefont {R.}~\bibnamefont {Tualle-Brouri}}, \bibinfo {author}
  {\bibfnamefont {S.~W.}\ \bibnamefont {McLaughlin}},\ and\ \bibinfo {author}
  {\bibfnamefont {P.}~\bibnamefont {Grangier}},\ }\bibfield  {title} {\bibinfo
  {title} {Quantum key distribution over
  $25\phantom{\rule{0.3em}{0ex}}\mathrm{km}$ with an all-fiber
  continuous-variable system},\ }\href@noop {} {\bibfield  {journal} {\bibinfo
  {journal} {Phys. Rev. A}\ }\textbf {\bibinfo {volume} {76}},\ \bibinfo
  {pages} {042305} (\bibinfo {year} {2007})}\BibitemShut {NoStop}%
\bibitem [{\citenamefont {Lo}\ \emph {et~al.}(2012)\citenamefont {Lo},
  \citenamefont {Curty},\ and\ \citenamefont {Qi}}]{lo2012measurement}%
  \BibitemOpen
  \bibfield  {author} {\bibinfo {author} {\bibfnamefont {H.-K.}\ \bibnamefont
  {Lo}}, \bibinfo {author} {\bibfnamefont {M.}~\bibnamefont {Curty}},\ and\
  \bibinfo {author} {\bibfnamefont {B.}~\bibnamefont {Qi}},\ }\bibfield
  {title} {\bibinfo {title} {Measurement-device-independent quantum key
  distribution},\ }\href@noop {} {\bibfield  {journal} {\bibinfo  {journal}
  {Phys. Rev. Lett.}\ }\textbf {\bibinfo {volume} {108}},\ \bibinfo {pages}
  {130503} (\bibinfo {year} {2012})}\BibitemShut {NoStop}%
\bibitem [{\citenamefont {Hillery}\ \emph {et~al.}(1999)\citenamefont
  {Hillery}, \citenamefont {Bu\ifmmode~\check{z}\else \v{z}\fi{}ek},\ and\
  \citenamefont {Berthiaume}}]{hillery1999quantum}%
  \BibitemOpen
  \bibfield  {author} {\bibinfo {author} {\bibfnamefont {M.}~\bibnamefont
  {Hillery}}, \bibinfo {author} {\bibfnamefont {V.}~\bibnamefont
  {Bu\ifmmode~\check{z}\else \v{z}\fi{}ek}},\ and\ \bibinfo {author}
  {\bibfnamefont {A.}~\bibnamefont {Berthiaume}},\ }\bibfield  {title}
  {\bibinfo {title} {Quantum secret sharing},\ }\href@noop {} {\bibfield
  {journal} {\bibinfo  {journal} {Phys. Rev. A}\ }\textbf {\bibinfo {volume}
  {59}},\ \bibinfo {pages} {1829} (\bibinfo {year} {1999})}\BibitemShut
  {NoStop}%
\bibitem [{\citenamefont {Fu}\ \emph {et~al.}(2015)\citenamefont {Fu},
  \citenamefont {Yin}, \citenamefont {Chen},\ and\ \citenamefont
  {Chen}}]{fu2015longdistance}%
  \BibitemOpen
  \bibfield  {author} {\bibinfo {author} {\bibfnamefont {Y.}~\bibnamefont
  {Fu}}, \bibinfo {author} {\bibfnamefont {H.-L.}\ \bibnamefont {Yin}},
  \bibinfo {author} {\bibfnamefont {T.-Y.}\ \bibnamefont {Chen}},\ and\
  \bibinfo {author} {\bibfnamefont {Z.-B.}\ \bibnamefont {Chen}},\ }\bibfield
  {title} {\bibinfo {title} {Long-distance measurement-device-independent
  multiparty quantum communication},\ }\href@noop {} {\bibfield  {journal}
  {\bibinfo  {journal} {Phys. Rev. Lett.}\ }\textbf {\bibinfo {volume} {114}},\
  \bibinfo {pages} {090501} (\bibinfo {year} {2015})}\BibitemShut {NoStop}%
\bibitem [{\citenamefont {Gu}\ \emph {et~al.}(2021{\natexlab{a}})\citenamefont
  {Gu}, \citenamefont {Xie}, \citenamefont {Liu}, \citenamefont {Fu},
  \citenamefont {Yin},\ and\ \citenamefont {Chen}}]{gu2021secure}%
  \BibitemOpen
  \bibfield  {author} {\bibinfo {author} {\bibfnamefont {J.}~\bibnamefont
  {Gu}}, \bibinfo {author} {\bibfnamefont {Y.-M.}\ \bibnamefont {Xie}},
  \bibinfo {author} {\bibfnamefont {W.-B.}\ \bibnamefont {Liu}}, \bibinfo
  {author} {\bibfnamefont {Y.}~\bibnamefont {Fu}}, \bibinfo {author}
  {\bibfnamefont {H.-L.}\ \bibnamefont {Yin}},\ and\ \bibinfo {author}
  {\bibfnamefont {Z.-B.}\ \bibnamefont {Chen}},\ }\bibfield  {title} {\bibinfo
  {title} {Secure quantum secret sharing without signal disturbance
  monitoring},\ }\href@noop {} {\bibfield  {journal} {\bibinfo  {journal} {Opt.
  Express}\ }\textbf {\bibinfo {volume} {29}},\ \bibinfo {pages} {32244}
  (\bibinfo {year} {2021}{\natexlab{a}})}\BibitemShut {NoStop}%
\bibitem [{\citenamefont {Schmid}\ \emph {et~al.}(2005)\citenamefont {Schmid},
  \citenamefont {Trojek}, \citenamefont {Bourennane}, \citenamefont
  {Kurtsiefer}, \citenamefont {{\.Z}ukowski},\ and\ \citenamefont
  {Weinfurter}}]{schmid2005experimental}%
  \BibitemOpen
  \bibfield  {author} {\bibinfo {author} {\bibfnamefont {C.}~\bibnamefont
  {Schmid}}, \bibinfo {author} {\bibfnamefont {P.}~\bibnamefont {Trojek}},
  \bibinfo {author} {\bibfnamefont {M.}~\bibnamefont {Bourennane}}, \bibinfo
  {author} {\bibfnamefont {C.}~\bibnamefont {Kurtsiefer}}, \bibinfo {author}
  {\bibfnamefont {M.}~\bibnamefont {{\.Z}ukowski}},\ and\ \bibinfo {author}
  {\bibfnamefont {H.}~\bibnamefont {Weinfurter}},\ }\bibfield  {title}
  {\bibinfo {title} {Experimental single qubit quantum secret sharing},\
  }\href@noop {} {\bibfield  {journal} {\bibinfo  {journal} {Phys. Rev. Lett.}\
  }\textbf {\bibinfo {volume} {95}},\ \bibinfo {pages} {230505} (\bibinfo
  {year} {2005})}\BibitemShut {NoStop}%
\bibitem [{\citenamefont {Gu}\ \emph {et~al.}(2021{\natexlab{b}})\citenamefont
  {Gu}, \citenamefont {Cao}, \citenamefont {Yin},\ and\ \citenamefont
  {Chen}}]{gu2021differential}%
  \BibitemOpen
  \bibfield  {author} {\bibinfo {author} {\bibfnamefont {J.}~\bibnamefont
  {Gu}}, \bibinfo {author} {\bibfnamefont {X.-Y.}\ \bibnamefont {Cao}},
  \bibinfo {author} {\bibfnamefont {H.-L.}\ \bibnamefont {Yin}},\ and\ \bibinfo
  {author} {\bibfnamefont {Z.-B.}\ \bibnamefont {Chen}},\ }\bibfield  {title}
  {\bibinfo {title} {Differential phase shift quantum secret sharing using a
  twin field},\ }\href@noop {} {\bibfield  {journal} {\bibinfo  {journal} {Opt.
  Express}\ }\textbf {\bibinfo {volume} {29}},\ \bibinfo {pages} {9165}
  (\bibinfo {year} {2021}{\natexlab{b}})}\BibitemShut {NoStop}%
\bibitem [{\citenamefont {Krawczyk}(1994)}]{krawczyk1994lfsr}%
  \BibitemOpen
  \bibfield  {author} {\bibinfo {author} {\bibfnamefont {H.}~\bibnamefont
  {Krawczyk}},\ }\bibfield  {title} {\bibinfo {title} {Lfsr-based hashing and
  authentication},\ }in\ \href@noop {} {\emph {\bibinfo {booktitle} {Annual
  International Cryptology Conference}}}\ (\bibinfo {year} {1994})\ pp.\
  \bibinfo {pages} {129--139}\BibitemShut {NoStop}%
\bibitem [{\citenamefont {Fung}\ \emph {et~al.}(2010)\citenamefont {Fung},
  \citenamefont {Ma},\ and\ \citenamefont {Chau}}]{fung2010practical}%
  \BibitemOpen
  \bibfield  {author} {\bibinfo {author} {\bibfnamefont {C.-H.~F.}\
  \bibnamefont {Fung}}, \bibinfo {author} {\bibfnamefont {X.}~\bibnamefont
  {Ma}},\ and\ \bibinfo {author} {\bibfnamefont {H.}~\bibnamefont {Chau}},\
  }\bibfield  {title} {\bibinfo {title} {Practical issues in
  quantum-key-distribution postprocessing},\ }\href@noop {} {\bibfield
  {journal} {\bibinfo  {journal} {Phys. Rev. A}\ }\textbf {\bibinfo {volume}
  {81}},\ \bibinfo {pages} {012318} (\bibinfo {year} {2010})}\BibitemShut
  {NoStop}%
\bibitem [{\citenamefont {Pompili}\ \emph {et~al.}(2021)\citenamefont
  {Pompili}, \citenamefont {Hermans}, \citenamefont {Baier}, \citenamefont
  {Beukers}, \citenamefont {Humphreys}, \citenamefont {Schouten}, \citenamefont
  {Vermeulen}, \citenamefont {Tiggelman}, \citenamefont {dos Santos~Martins},
  \citenamefont {Dirkse}, \citenamefont {Wehner},\ and\ \citenamefont
  {Hanson}}]{Pompili2021network}%
  \BibitemOpen
  \bibfield  {author} {\bibinfo {author} {\bibfnamefont {M.}~\bibnamefont
  {Pompili}}, \bibinfo {author} {\bibfnamefont {S.~L.~N.}\ \bibnamefont
  {Hermans}}, \bibinfo {author} {\bibfnamefont {S.}~\bibnamefont {Baier}},
  \bibinfo {author} {\bibfnamefont {H.~K.~C.}\ \bibnamefont {Beukers}},
  \bibinfo {author} {\bibfnamefont {P.~C.}\ \bibnamefont {Humphreys}}, \bibinfo
  {author} {\bibfnamefont {R.~N.}\ \bibnamefont {Schouten}}, \bibinfo {author}
  {\bibfnamefont {R.~F.~L.}\ \bibnamefont {Vermeulen}}, \bibinfo {author}
  {\bibfnamefont {M.~J.}\ \bibnamefont {Tiggelman}}, \bibinfo {author}
  {\bibfnamefont {L.}~\bibnamefont {dos Santos~Martins}}, \bibinfo {author}
  {\bibfnamefont {B.}~\bibnamefont {Dirkse}}, \bibinfo {author} {\bibfnamefont
  {S.}~\bibnamefont {Wehner}},\ and\ \bibinfo {author} {\bibfnamefont
  {R.}~\bibnamefont {Hanson}},\ }\bibfield  {title} {\bibinfo {title}
  {Realization of a multinode quantum network of remote solid-state qubits},\
  }\href@noop {} {\bibfield  {journal} {\bibinfo  {journal} {Science}\ }\textbf
  {\bibinfo {volume} {372}},\ \bibinfo {pages} {259} (\bibinfo {year}
  {2021})}\BibitemShut {NoStop}%
\bibitem [{\citenamefont {Yin}\ \emph {et~al.}(2020{\natexlab{a}})\citenamefont
  {Yin}, \citenamefont {Liu}, \citenamefont {Dai}, \citenamefont {Ci},
  \citenamefont {Gu}, \citenamefont {Gao}, \citenamefont {Wang},\ and\
  \citenamefont {Shen}}]{yin2020experimental}%
  \BibitemOpen
  \bibfield  {author} {\bibinfo {author} {\bibfnamefont {H.-L.}\ \bibnamefont
  {Yin}}, \bibinfo {author} {\bibfnamefont {P.}~\bibnamefont {Liu}}, \bibinfo
  {author} {\bibfnamefont {W.-W.}\ \bibnamefont {Dai}}, \bibinfo {author}
  {\bibfnamefont {Z.-H.}\ \bibnamefont {Ci}}, \bibinfo {author} {\bibfnamefont
  {J.}~\bibnamefont {Gu}}, \bibinfo {author} {\bibfnamefont {T.}~\bibnamefont
  {Gao}}, \bibinfo {author} {\bibfnamefont {Q.-W.}\ \bibnamefont {Wang}},\ and\
  \bibinfo {author} {\bibfnamefont {Z.-Y.}\ \bibnamefont {Shen}},\ }\bibfield
  {title} {\bibinfo {title} {Experimental composable security decoy-state
  quantum key distribution using time-phase encoding},\ }\href@noop {}
  {\bibfield  {journal} {\bibinfo  {journal} {Opt. Express}\ }\textbf {\bibinfo
  {volume} {28}},\ \bibinfo {pages} {29479} (\bibinfo {year}
  {2020}{\natexlab{a}})}\BibitemShut {NoStop}%
\bibitem [{\citenamefont {Long}\ \emph {et~al.}(2022)\citenamefont {Long},
  \citenamefont {Pan}, \citenamefont {Sheng}, \citenamefont {Xue},
  \citenamefont {Lu},\ and\ \citenamefont {Hanzo}}]{long2022evolutionary}%
  \BibitemOpen
  \bibfield  {author} {\bibinfo {author} {\bibfnamefont {G.-L.}\ \bibnamefont
  {Long}}, \bibinfo {author} {\bibfnamefont {D.}~\bibnamefont {Pan}}, \bibinfo
  {author} {\bibfnamefont {Y.-B.}\ \bibnamefont {Sheng}}, \bibinfo {author}
  {\bibfnamefont {Q.}~\bibnamefont {Xue}}, \bibinfo {author} {\bibfnamefont
  {J.}~\bibnamefont {Lu}},\ and\ \bibinfo {author} {\bibfnamefont
  {L.}~\bibnamefont {Hanzo}},\ }\bibfield  {title} {\bibinfo {title} {An
  evolutionary pathway for the quantum internet relying on secure classical
  repeaters},\ }\href@noop {} {\bibfield  {journal} {\bibinfo  {journal} {arXiv
  preprint arXiv:2202.03619}\ } (\bibinfo {year} {2022})}\BibitemShut {NoStop}%
\bibitem [{\citenamefont {Schneier}(2015)}]{schneier2015secrets}%
  \BibitemOpen
  \bibfield  {author} {\bibinfo {author} {\bibfnamefont {B.}~\bibnamefont
  {Schneier}},\ }\href@noop {} {\emph {\bibinfo {title} {Secrets and lies:
  digital security in a networked world}}}\ (\bibinfo  {publisher} {John Wiley
  \& Sons},\ \bibinfo {year} {2015})\BibitemShut {NoStop}%
\bibitem [{\citenamefont {Rabin}(1980)}]{rabin1980probabilistic}%
  \BibitemOpen
  \bibfield  {author} {\bibinfo {author} {\bibfnamefont {M.~O.}\ \bibnamefont
  {Rabin}},\ }\bibfield  {title} {\bibinfo {title} {Probabilistic algorithms in
  finite fields},\ }\href@noop {} {\bibfield  {journal} {\bibinfo  {journal}
  {SIAM Journal on computing}\ }\textbf {\bibinfo {volume} {9}},\ \bibinfo
  {pages} {273} (\bibinfo {year} {1980})}\BibitemShut {NoStop}%
\bibitem [{\citenamefont {Von Zur~Gathen}\ and\ \citenamefont
  {Gerhard}(2013)}]{von2013modern}%
  \BibitemOpen
  \bibfield  {author} {\bibinfo {author} {\bibfnamefont {J.}~\bibnamefont {Von
  Zur~Gathen}}\ and\ \bibinfo {author} {\bibfnamefont {J.}~\bibnamefont
  {Gerhard}},\ }\href@noop {} {\emph {\bibinfo {title} {Modern computer
  algebra}}}\ (\bibinfo  {publisher} {Cambridge university press},\ \bibinfo
  {year} {2013})\BibitemShut {NoStop}%
\bibitem [{\citenamefont {Zeng}\ \emph {et~al.}(2020)\citenamefont {Zeng},
  \citenamefont {Wu},\ and\ \citenamefont {Ma}}]{zeng2020symmetry}%
  \BibitemOpen
  \bibfield  {author} {\bibinfo {author} {\bibfnamefont {P.}~\bibnamefont
  {Zeng}}, \bibinfo {author} {\bibfnamefont {W.}~\bibnamefont {Wu}},\ and\
  \bibinfo {author} {\bibfnamefont {X.}~\bibnamefont {Ma}},\ }\bibfield
  {title} {\bibinfo {title} {Symmetry-protected privacy: beating the
  rate-distance linear bound over a noisy channel},\ }\href@noop {} {\bibfield
  {journal} {\bibinfo  {journal} {Phys. Rev. Appl.}\ }\textbf {\bibinfo
  {volume} {13}},\ \bibinfo {pages} {064013} (\bibinfo {year}
  {2020})}\BibitemShut {NoStop}%
\bibitem [{\citenamefont {Comandar}\ \emph {et~al.}(2016)\citenamefont
  {Comandar}, \citenamefont {Lucamarini}, \citenamefont {Fr{\"o}hlich},
  \citenamefont {Dynes}, \citenamefont {Sharpe}, \citenamefont {Tam},
  \citenamefont {Yuan}, \citenamefont {Penty},\ and\ \citenamefont
  {Shields}}]{comandar2016quantum}%
  \BibitemOpen
  \bibfield  {author} {\bibinfo {author} {\bibfnamefont {L.}~\bibnamefont
  {Comandar}}, \bibinfo {author} {\bibfnamefont {M.}~\bibnamefont
  {Lucamarini}}, \bibinfo {author} {\bibfnamefont {B.}~\bibnamefont
  {Fr{\"o}hlich}}, \bibinfo {author} {\bibfnamefont {J.}~\bibnamefont {Dynes}},
  \bibinfo {author} {\bibfnamefont {A.}~\bibnamefont {Sharpe}}, \bibinfo
  {author} {\bibfnamefont {S.-B.}\ \bibnamefont {Tam}}, \bibinfo {author}
  {\bibfnamefont {Z.}~\bibnamefont {Yuan}}, \bibinfo {author} {\bibfnamefont
  {R.}~\bibnamefont {Penty}},\ and\ \bibinfo {author} {\bibfnamefont
  {A.}~\bibnamefont {Shields}},\ }\bibfield  {title} {\bibinfo {title} {Quantum
  key distribution without detector vulnerabilities using optically seeded
  lasers},\ }\href@noop {} {\bibfield  {journal} {\bibinfo  {journal} {Nat.
  Photonics}\ }\textbf {\bibinfo {volume} {10}},\ \bibinfo {pages} {312}
  (\bibinfo {year} {2016})}\BibitemShut {NoStop}%
\bibitem [{\citenamefont {Wang}(2005)}]{Wang2005Beating}%
  \BibitemOpen
  \bibfield  {author} {\bibinfo {author} {\bibfnamefont {X.-B.}\ \bibnamefont
  {Wang}},\ }\bibfield  {title} {\bibinfo {title} {Beating the
  photon-number-splitting attack in practical quantum cryptography},\
  }\href@noop {} {\bibfield  {journal} {\bibinfo  {journal} {Phys. Rev. Lett.}\
  }\textbf {\bibinfo {volume} {94}},\ \bibinfo {pages} {230503} (\bibinfo
  {year} {2005})}\BibitemShut {NoStop}%
\bibitem [{\citenamefont {Lo}\ \emph {et~al.}(2005)\citenamefont {Lo},
  \citenamefont {Ma},\ and\ \citenamefont {Chen}}]{Lo:2005:Decoy}%
  \BibitemOpen
  \bibfield  {author} {\bibinfo {author} {\bibfnamefont {H.-K.}\ \bibnamefont
  {Lo}}, \bibinfo {author} {\bibfnamefont {X.}~\bibnamefont {Ma}},\ and\
  \bibinfo {author} {\bibfnamefont {K.}~\bibnamefont {Chen}},\ }\bibfield
  {title} {\bibinfo {title} {Decoy state quantum key distribution},\
  }\href@noop {} {\bibfield  {journal} {\bibinfo  {journal} {Phys. Rev. Lett.}\
  }\textbf {\bibinfo {volume} {94}},\ \bibinfo {pages} {230504} (\bibinfo
  {year} {2005})}\BibitemShut {NoStop}%
\bibitem [{\citenamefont {Yu}\ \emph {et~al.}(2016)\citenamefont {Yu},
  \citenamefont {Zhou},\ and\ \citenamefont {Wang}}]{Yu:2016:Reexamination}%
  \BibitemOpen
  \bibfield  {author} {\bibinfo {author} {\bibfnamefont {Z.-W.}\ \bibnamefont
  {Yu}}, \bibinfo {author} {\bibfnamefont {Y.-H.}\ \bibnamefont {Zhou}},\ and\
  \bibinfo {author} {\bibfnamefont {X.-B.}\ \bibnamefont {Wang}},\ }\bibfield
  {title} {\bibinfo {title} {Reexamination of decoy-state quantum key
  distribution with biased bases},\ }\href@noop {} {\bibfield  {journal}
  {\bibinfo  {journal} {Phys. Rev. A}\ }\textbf {\bibinfo {volume} {93}},\
  \bibinfo {pages} {032307} (\bibinfo {year} {2016})}\BibitemShut {NoStop}%
\bibitem [{\citenamefont {Lim}\ \emph {et~al.}(2014)\citenamefont {Lim},
  \citenamefont {Curty}, \citenamefont {Walenta}, \citenamefont {Xu},\ and\
  \citenamefont {Zbinden}}]{lim2014concise}%
  \BibitemOpen
  \bibfield  {author} {\bibinfo {author} {\bibfnamefont {C.~C.~W.}\
  \bibnamefont {Lim}}, \bibinfo {author} {\bibfnamefont {M.}~\bibnamefont
  {Curty}}, \bibinfo {author} {\bibfnamefont {N.}~\bibnamefont {Walenta}},
  \bibinfo {author} {\bibfnamefont {F.}~\bibnamefont {Xu}},\ and\ \bibinfo
  {author} {\bibfnamefont {H.}~\bibnamefont {Zbinden}},\ }\bibfield  {title}
  {\bibinfo {title} {Concise security bounds for practical decoy-state quantum
  key distribution},\ }\href@noop {} {\bibfield  {journal} {\bibinfo  {journal}
  {Phys. Rev. A}\ }\textbf {\bibinfo {volume} {89}},\ \bibinfo {pages} {022307}
  (\bibinfo {year} {2014})}\BibitemShut {NoStop}%
\bibitem [{\citenamefont {Yin}\ \emph {et~al.}(2020{\natexlab{b}})\citenamefont
  {Yin}, \citenamefont {Zhou}, \citenamefont {Gu}, \citenamefont {Xie},
  \citenamefont {Lu},\ and\ \citenamefont {Chen}}]{yin2020tight}%
  \BibitemOpen
  \bibfield  {author} {\bibinfo {author} {\bibfnamefont {H.-L.}\ \bibnamefont
  {Yin}}, \bibinfo {author} {\bibfnamefont {M.-G.}\ \bibnamefont {Zhou}},
  \bibinfo {author} {\bibfnamefont {J.}~\bibnamefont {Gu}}, \bibinfo {author}
  {\bibfnamefont {Y.-M.}\ \bibnamefont {Xie}}, \bibinfo {author} {\bibfnamefont
  {Y.-S.}\ \bibnamefont {Lu}},\ and\ \bibinfo {author} {\bibfnamefont {Z.-B.}\
  \bibnamefont {Chen}},\ }\bibfield  {title} {\bibinfo {title} {Tight security
  bounds for decoy-state quantum key distribution},\ }\href@noop {} {\bibfield
  {journal} {\bibinfo  {journal} {Sci. Rep.}\ }\textbf {\bibinfo {volume}
  {10}},\ \bibinfo {pages} {14312} (\bibinfo {year}
  {2020}{\natexlab{b}})}\BibitemShut {NoStop}%
\end{thebibliography}
%

\end{document}